%% file: main.tex
\documentclass[sigconf]{acmart}

\usepackage{enumitem}
\usepackage{subcaption}
\usepackage{makecell}
\usepackage{multirow}

\copyrightyear{2024}
\acmYear{2024}
\setcopyright{rightsretained}
\acmConference[CHI '24]{Proceedings of the CHI Conference on Human Factors in Computing Systems}{May 11--16, 2024}{Honolulu, HI, USA}
\acmBooktitle{Proceedings of the CHI Conference on Human Factors in Computing Systems (CHI '24), May 11--16, 2024, Honolulu, HI, USA}
\acmDOI{10.1145/3613904.3641969}
\acmISBN{979-8-4007-0330-0/24/05}

\begin{document}

\title[Supporting Business Document Workflows via Collection-Centric Information Foraging]{Supporting Business Document Workflows via Collection-Centric Information Foraging with Large Language Models}

\author{Raymond Fok}
\email{rayfok@cs.washington.edu}
\affiliation{
  \institution{University of Washington}
  \city{Seattle}
  \state{WA}
  \country{USA}
}
\authornote{Work completed during an internship at Adobe Research.}

\author{Nedim Lipka}
\email{lipka@adobe.com}
\affiliation{
  \institution{Adobe Research}
  \city{San Jose}
  \state{CA}
  \country{USA}
}

\author{Tong Sun}
\email{tsun@adobe.com}
\affiliation{
  \institution{Adobe Research}
  \city{San Jose}
  \state{CA}
  \country{USA}
}

\author{Alexa Siu}
\email{asiu@adobe.com}
\affiliation{
  \institution{Adobe Research}
  \city{San Jose}
  \state{CA}
  \country{USA}
}


\include{macros}

\begin{abstract}
\input{sections/00_abstract}
\end{abstract}



\maketitle

\input{sections/01_introduction}
\input{sections/02_related_work}
\input{sections/03_formative}
\input{sections/04_system}
\input{sections/05_evaluation}
\input{sections/06_discussion}
\input{sections/07_conclusion}

\begin{acks}
    The authors thank Joe Barrow and Jack Wang for their helpful feedback on the initial system design. We also thank all participants across our numerous user studies, without whom this work would not have been possible.
\end{acks}

\bibliographystyle{ACM-Reference-Format}
\bibliography{combined}

\input{sections/08_appendix}

\end{document}

%% file: macros.tex
\newcommand\todo[1]{\textcolor{red}{[TODO]: #1}}
\newcommand\ray[1]{\textcolor{blue}{[Ray]: #1}}
\newcommand\alexa[1]{\textcolor{red}{[Alexa]: #1}}

\newcommand\edit[1]{{\leavevmode\color{red} #1}}

\newcommand\system{\textsc{Marco}}

\newcolumntype{L}[1]{>{\raggedright\let\newline\\\arraybackslash\hspace{0pt}}m{#1}}
\newcolumntype{C}[1]{>{\centering\let\newline\\\arraybackslash\hspace{0pt}}m{#1}}
\newcolumntype{R}[1]{>{\raggedleft\let\newline\\\arraybackslash\hspace{0pt}}m{#1}}

%% file: sections/00_abstract.tex
Knowledge workers often need to extract and analyze information from a collection of documents to solve complex information tasks in the workplace, e.g., hiring managers reviewing resumes or analysts assessing risk in contracts. However, foraging for relevant information can become tedious and repetitive over many documents and criteria of interest. We introduce \system{}, a mixed-initiative workspace supporting sensemaking over diverse business document collections. Through collection-centric assistance, \system{} reduces the cognitive costs of extracting and structuring information, allowing users to prioritize comparative synthesis and decision making processes. Users interactively communicate their information needs to an AI assistant using natural language and compose schemas that provide an overview of a document collection. Findings from a usability study (n=16) demonstrate that when using \system{}, users complete sensemaking tasks 16\% more quickly, with less effort, and without diminishing accuracy. A design probe with seven domain experts identifies how \system{} can benefit various real-world workflows.

%% file: sections/01_introduction.tex
\section{Introduction}

\input{figures/ui-overview}

Knowledge workers derive insights from information to accomplish complex information tasks, with the intention to solve problems, plan actions, and make decisions. For instance, a business analyst may want to determine the best negotiation strategy given related business contracts, a hiring manager may want to select candidates from a pool of resumes, or a researcher may want to conduct a literature review across multiple scholarly articles. In service of their goals, people need to triage, search through, and make sense of copious information throughout their document collections.
These goal-driven processes can be viewed within a sensemaking framework (Figure~\ref{fig:sensemaking}), consisting of two interconnected loops of foraging and sensemaking activities ~\cite{pirolli_information_1995, pirolli_sensemaking_2005}.
Prior work has studied how technology can support this sensemaking process within domains such as exploratory online research~\cite{chang_searchlens_2019, ramos_forsense_2022, kuznetsov_fuse_2022, chang_mesh_2020} or scholarly synthesis~\cite{kang_synergi_2023, kang_threddy_2022, zhang_citesense_2008}. But one area that has received less attention is document-centered assistance within the workplace~\cite{jahanbakhsh_understanding_2022}.

To address this gap, we investigate the potential for digital assistance to support information foraging and sensemaking over business document collections.
We first conducted formative interviews with 12 knowledge workers across diverse business functions, seeking to understand the current tools, strategies, and pain points within their workflows.
We found that despite working with a diversity of documents, e.g., legal or financial contracts, and exhibiting a wide range of goals, participants expressed a common challenge of searching for information across their documents.
Current foraging processes were described as tedious and time-consuming, involving manually searching for, extracting, and organizing information into structured representations such as a spreadsheet, before repeating for all subsequent documents.
Our findings revealed that participants' predominant pain points focused on information foraging, which often comprised the bulk of their workflow, despite being an intermediary albeit critical step in service of their own specialized sensemaking goals, such as to derive insights or inform decision making.
Given the manual nature of foraging processes and the lack of supportive tools, many participants were optimistic about the potential for AI assistance to complement their workflows.

In this work, we propose a collection-centric interaction paradigm in which knowledge workers primarily engage with their documents as a cohesive collection rather than as discrete documents.
We reify this vision within a novel interactive system, \system{}, that leverages AI assistance to help knowledge workers forage for similar information across many documents, organize gathered information, and synthesize collection-level insights. \system{} presents a workspace integrating three views: Notebook View, Table View, and Document View (Figure~\ref{fig:ui-overview}). Within \system{}, users build up a notebook of cells containing rich text serving as a note-taking space and \textit{actions} executed over a document collection (Notebook View).



Actions in Marco allow users to delegate foraging tasks to AI assistance, such that users can instead prioritize their focus on sensemaking tasks (Figure~\ref{fig:sensemaking}).
For instance, users can perform a lexical or semantic search across each document in their collection in parallel with a \texttt{Search} action, ask questions of their documents with an \texttt{Ask} action, or summarize documents along desired dimensions with a \texttt{Summarize} action. Actions can be executed over a single document, multiple documents, or by default, the entire collection. Foraged information is organized into a tabular schema, allowing users to inspect specific details within documents and also compare across documents. As actions are executed, \system{} joins the foraged information into an aggregate table (Table View), aiding sensemaking. Finally, \system{} actively recommends actions tailored to users' specific document collections, defined goals, and past actions, to kickstart or encourage future foraging directions.

Through a controlled usability study and a design probe with domain experts, we sought to answer three research questions:
\begin{enumerate}[label=\textit{RQ\arabic*.}]
    \item How does \system{} impact users' performance and experience sensemaking over business document collections?
    \item How do knowledge workers make use of \system{}'s features when working with business document collections?
    \item How do knowledge workers perceive and interact with imprecise AI assistance in \system{}?
\end{enumerate}

Findings from our usability study demonstrate the efficacy of \system{} when compared to a baseline approach. We found participants completed tasks 16\% more quickly using \system{}, and self-reported information was easier to find and required less effort to synthesize, with no difference in confidence or accuracy. Interaction logs and qualitative analysis complemented these findings, highlighting how \system{}'s suite of actions reduced tedium in information foraging processes, supported collection-level analyses, and encouraged verification of imperfect AI assistance. Our subsequent design probe identified opportunities for \system{} to accelerate sensemaking within real-world business workflows and suggested additional considerations for supporting users in reconciling with imprecise AI assistance. 

In summary, this paper contributes:
\begin{itemize}
\item Insights from a formative study highlighting challenges people face when working with collections of business documents and four design goals that emerged to inform intelligent sensemaking support tools for business workflows.
\item \system{}, a novel interactive workspace with a suite of natural language, document-centered \textit{actions} that facilitate foraging and sensemaking over business document collections.
\item Findings from a usability study (n=16) showing \system{} improves efficiency and reduces effort in information-seeking tasks over document collections, and insights from a design probe with domain experts (n=7) suggesting how such support can benefit real-world business document workflows.
\end{itemize}

%% file: figures/ui-overview.tex
\begin{figure*}[t]
    \centering
    \includegraphics[width=0.95\textwidth]{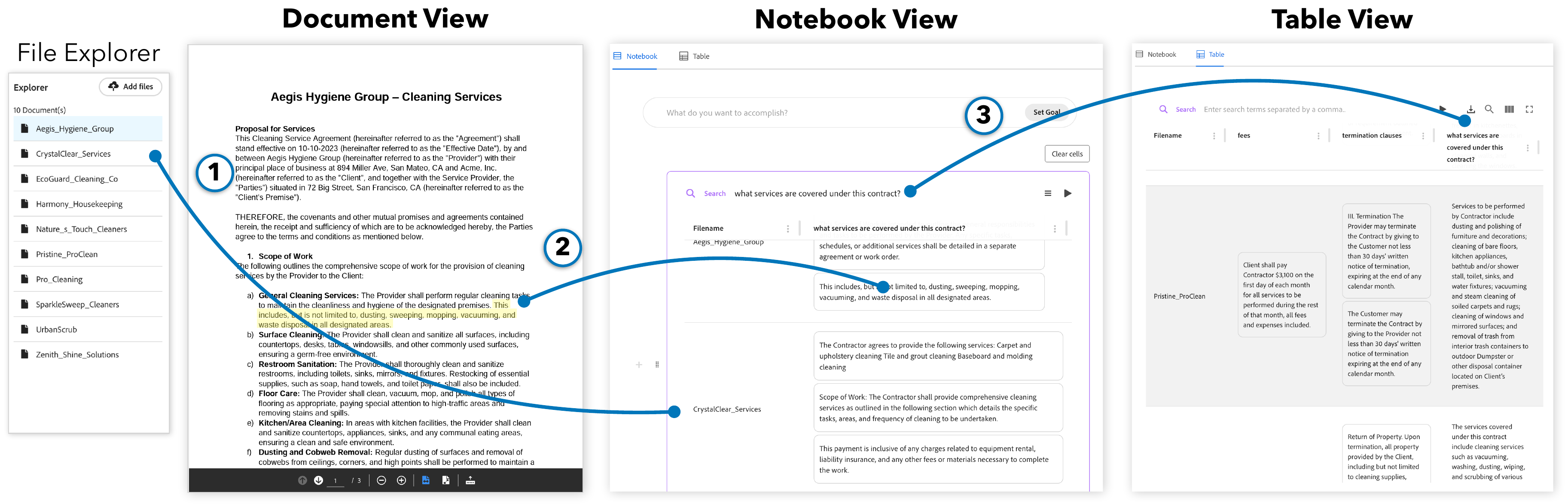}
    \caption{\system{} is a mixed-initiative workspace for sensemaking over document collections. \system{} integrates three views: a \textbf{Document View} renders a PDF document, a \textbf{Notebook View} provides document-centered \textit{actions} leveraging LLMs, and a \textbf{Table View} provides a collection-level overview. Actions in the Notebook View encode relevant information within result tables, with one row per document (1). Responses can be verified with in-context highlights within the Document View (2). Results across actions are concatenated into a Table View to support collection-level analysis (3).}
    \Description[Three interface screenshots]{Three screenshots of an interface side-by-side. The first shows a PDF document, the second a notebook with a purple box, and the third a table with two rows and three columns. Three numbered blue lines connect each adjacent interface.}
    \label{fig:ui-overview}
\end{figure*}

%% file: sections/02_related_work.tex
\section{Related Work}

\input{figures/sensemaking}

\subsection{Sensemaking and Sensemaking Support}
Our work draws inspiration from information foraging and sensemaking theories which characterize how people extract and organize information in service of their goals~\cite{pirolli_information_1995, pirolli_sensemaking_2005, russell_cost_1993}. The analysis of information, or sensemaking---underlying processes such as reasoning, problem solving, and decision making---can be organized into two major loops of activity: \textit{foraging}, in which people extract relevant information from a broader corpus, and \textit{sensemaking}, in which people reason about the expected utility of the foraged information and iteratively refine organizational structures, or \textit{schemas}, that encode the information to inform deeper synthesis (Figure~\ref{fig:sensemaking}).

Early sensemaking support focused on helping people explore collections of online web pages. For example, \textit{Scatter/Gather} uses clustering to browse large online collections~\cite{cutting_scattergather_1992}, \textit{SenseMaker} focuses on foraging across heterogeneous web pages~\cite{baldonado_sensemaker_1997}, and faceted search techniques help filter online content~\cite{hearst_nlp_2009, english_hierarchical_2002, schraefel_mspace_2006}. Some work has introduced approaches to improve the foraging process by reducing the costs of finding and saving relevant information~\cite{kittur_costs_2013, ramos_forsense_2022, schraefel_hunter_2002, dontcheva_relations_2007, liu_wigglite_2022}, while others have focused on assisting in the creation of schemas to encode the foraged information~\cite{chang_mesh_2020, chang_searchlens_2019, chen_explaining_2017, dontcheva_relations_2007, chen_experiment_2014}.

The most benefit to sensemaking has arguably been observed in systems that more closely integrate foraging and sensemaking loops, for example within exploratory online search~\cite{nguyen_sensemap_2016, hearst_sewing_2013, hahn_bento_2018, chang_searchlens_2019, ramos_forsense_2022, suh_sensecape_2023}, programming~\cite{liu_unakite_2019, liu_reuse_2021, liu_crystalline_2022}, mobile information exploration~\cite{swearngin_scraps_2021}, or scholarly literature review~\cite{zhang_citesense_2008, kang_threddy_2022, palani_relatedly_2023}.
In our work, we contribute to this rich thread of research, but focus on the context of sensemaking over business documents in the workplace, an area in which document-centered assistance has seen limited progress. Furthermore, we demonstrate a novel means of integrating the foraging and sensemaking loops. By automatically ``mapping'' users' queries across many documents, \system{} both helps reduce the costs of foraging and creates the organizational structures that facilitate cross-document comparison. Closing the loop, \system{} then uses these structures to suggest subsequent foraging actions.

\subsection{Document Consumption in the Workplace}
Knowledge work can consist of a variety of professions and work-oriented goals. An underlying characteristic of these workers is high expertise in tacit or declarative knowledge (knowing the facts of the work) and procedural knowledge (knowing how to do the work)~\cite{heidary2011activity}.
In the space of document processes, several works have sought to understand knowledge workers’ practices~\cite{adler1998diary, russell_cost_1993}. An in-depth diary study with knowledge workers from diverse functions estimated document-related activities accounted for 82\% of users’ working time~\cite{adler1998diary}. Russell et al. observed that data extraction and encoding from documents accounted for 75\% of analysts’ time when working with document collections~\cite{russell_cost_1993}. 
A more recent survey reported knowledge workers spend anywhere from 1-3 hours simply attempting to locate information or a specific document~\cite{zapier_office}.

Toward addressing these challenges, Jahanbakshsh et al. characterized the various types of document-centric assistance that could support knowledge workers. They found that needs over documents varied for different document types, and included factual, reasoning, and overview types of questions~\cite{jahanbakhsh_understanding_2022}.
Recent work has started to identify opportunities for AI assistance in document processes, for instance through Q\&A systems~\cite{jahanbakhsh_understanding_2022, ter_hoeve_conversations_2020} and general-purpose LLM-powered tools~\cite{cambon2023early, dell2023navigating}. However, these works indicate that adoption and use of automation tools for document-centered assistance have remained limited~\cite{jahanbakhsh_understanding_2022}. 
Moreover, a case study with knowledge workers and AI assistants reported a learning curve for users to teach AI assistance their tacit knowledge about their working context~\cite{ferreira2019should}.
\system{} aims to support knowledge workers’ sensemaking activities over document collections by leveraging their domain expertise to complement AI capabilities. 

Prior work has explored interactions to support document-centric tasks focusing on specific types of documents, such as legal documents~\cite{han2020textlets, roegiest2018redesigning}, scientific documents~\cite{lo2023semantic, matejka2021paperForager, palani_relatedly_2023, han_passages_2022}, humanities studies~\cite{moretti2016alcide}, and patents~\cite{hui2005extracting, han_passages_2022}. These systems make use of common characteristics present in these document types to support domain-specific search and analysis tasks. For instance, \textit{Passages} supports scientists and patent examiners in managing document provenance and organizing relevant text selections across documents in one view~\cite{han_passages_2022}. \textit{PaperForager} helps users conduct a literature review by reducing the cost of transitioning between browsing a collection to reading an individual page of interest~\cite{matejka2021paperForager}. In our work, we aim to understand the needs of knowledge workers in different business-related domains and equip \system{} with general-purpose tools that support complex information tasks and analyses over a variety of document types.

\subsection{Human-LLM Interaction} 
Large language models (LLMs) have engendered myriad applications to support sensemaking, e.g., within online research~\cite{suh_sensecape_2023}, scholarly research~\cite{kang_synergi_2023}, and argumentative writing~\cite{zhang_visar_2023}. In conversational applications such as ChatGPT~\cite{chatgpt} and Bard~\cite{bard}, LLMs have demonstrated impressive capabilities in answering users' open-domain questions, and be further refined to answer questions given user-provided documents. However, unlike question answering over individual documents~\cite{zhao_talk_2020}, the non-linear and dynamic workflows of sensemaking motivate exploration of a novel design space of human-LLM interactions. Recent work has shown LLM-powered applications can go beyond a linear, chat-like interaction paradigm, for instance by transposing text-based responses into flexible graphical representations~\cite{jiang_graphologue_2023}, enabling recursive summarization~\cite{kang_synergi_2023}, or supporting multilevel exploration of information~\cite{suh_sensecape_2023}. 

Through \system{}, we envision how LLMs can be used to support flexible sensemaking processes over document collections. \system{} draws on numerous LLM capabilities, from information extraction for semantic search and multi-document question answering, to recommendations for follow-up foraging directions. Despite their success, one well-known challenge of LLMs is their tendency to hallucinate~\cite{shen_chatgpt_2023,bang_multitask_2023, mundler_self-contradictory_2023, ji_survey_2023}. In the context of knowledge work, users therefore have to calibrate their trust in these models, such as through manual verification of the models' responses.

Guiding principles for mixed-initiative user interfaces in which ``intelligent services and users may often collaborate efficiently to achieve the user’s goals'' were proposed over two decades ago by \citet{horvitz_principles_1999}, and later modernized into 18 design guidelines for human-AI interaction in AI-infused systems~\cite{amershi_guidelines_2019}. We lean on many of these guidelines to build interactions within \system{} which adapt to user context (\textit{``Show contextually relevant information''}), reduce friction in collaboration between users and intelligent agents (\textit{``Support efficient invocation and dismissal''}), recover from imperfect AI systems (\textit{``Support efficient correction of errors''}), and learn from users' interactions over time (\textit{``Learn from user behavior''}). Through the design of \system{}, we offer an initial vision of how these principles of human-AI interaction can be adapted to mixed-initiative, LLM-powered systems for document assistance.

%% file: figures/sensemaking.tex
\begin{figure*}[t]
    \centering
    \includegraphics[width=0.95\textwidth]{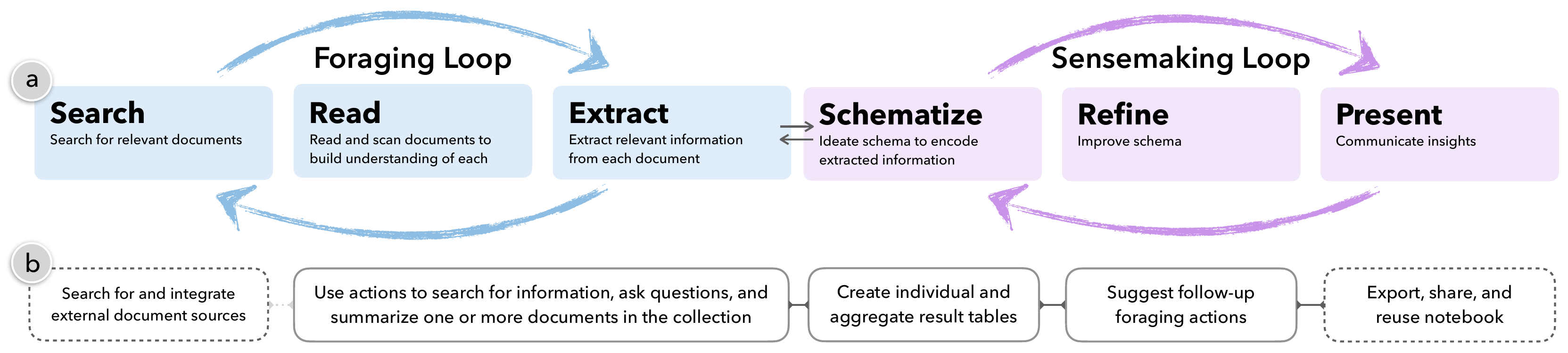}
    \caption{
        (a) The sensemaking process consists of two iterative loops of activity: information foraging and sensemaking~\cite{pirolli_sensemaking_2005}. (b) \system{} was designed to support different stages of the sensemaking process  in business workflows. Using natural language, users delegate foraging tasks to AI assistance (foraging loop), enabling users to focus on verifying AI responses, refining information schemas, and synthesizing information (sensemaking loop). Solid lines indicate capabilities available in \system{}.
    }
    \Description [A block diagram of the sensemaking process]{A block diagram of the general sensemaking process, and text describing how Marco implements aspects of it. Six blocks side-by-side, with the text Search, Read, Extract, Schematize, Refine, and Present. The first three blocks are blue and connected, labeled as the foraging loop. The second three blocks are purple and connected, labeled as the sensemaking loop. Boxes below each of the blocks describe how Marco supports the processes in each block.}
    \label{fig:sensemaking}
\end{figure*}

%% file: sections/03_formative.tex
\section{Understanding Business Document Collection Workflows}
\label{sec:formative-study}
To better understand the challenges people encounter within current document-centered business workflows, we conducted a formative interview study with knowledge workers across various functional areas of business.

\subsection{Participants}
We recruited 12 participants from within a large software organization using purposive sampling (Table~\ref{tab:formative-study-overview}). Participants spanned diverse functional areas, including finance, procurement, legal, and management, and were employed in sectors across technology, education, and healthcare. All participants were personally responsible for or managed teams whose responsibilities involved reviewing large collections of documents. One participant had less than 5 years of experience, one had 5--10 years of experience, six had 11--20 years of experience, and four had more than 20 years of experience performing tasks involving business documents. Participants were thanked for their time but not compensated.

\subsection{Procedure and Analysis}
We conducted semi-structured interviews, asking participants about the primary tasks they conducted for their role, the types and volume of documents they typically review, their strategies and goals while reviewing, and their current usage or perception of AI assistance for their tasks. Interviews were conducted remotely, recorded, and transcribed. One author went through the transcripts and coded them for themes using an open thematic analysis process~\cite{braun2006using, boyatzis1998transforming}. The research team then discussed and iterated upon the themes, informing a set of clear design goals.

\input{tables/formative_study_participants}

\subsection{Findings} \label{sec:formative-study-findings}
Participants described various document collections, document-centric goals, and information foraging and organizational strategies when working with business documents. Despite this diversity, all participants highlighted a common challenge in how their current workflows lacked the appropriate tools to support repetitive and tedious information foraging needs over many documents. Below, we highlight the main findings of our study.

\textbf{Knowledge workers rely on structured representations to facilitate consistent information foraging and enable comparison across documents.}
    Participants often mentioned workflows in which they sought to synthesize information across multiple business documents to inform decision making.
    For instance, one procurement specialist who reviewed collections of vendor contracts to optimize future negotiations described needing to \textit{``extract information, consolidate, and then make some meaning out of it''} (P9).
    Another financial planning analyst who compiled reports for executives by summarizing patterns from previous earnings calls characterized their workflow as ``evaluating a whole bunch of documents \dots connecting the dots'' (P1).
    To reason across documents, participants needed to effectively review and compare documents along multiple dimensions of interest.
    However, performing this comparison was challenging, as business documents---even those within the same collection, e.g., a set of legal contracts---can vary in length, structure, and content.
    Current reviews of multiple documents are typically conducted by numerous human analysts across a functional team, and therefore what information is extracted from each document and how it is organized can differ.
    Multiple rounds of review might be required for certain tasks, as one analyst described: \textit{``We have an analyst go in and review that contract, especially paying close attention to any non-standard terms or conditions. Then depending on the dollar value, there might be a second review, either by a peer or by a manager''} (P8).
    To encourage consistency across documents and people, participants mentioned strategies for guiding their reviews: using a pre-defined form or checklist (P7, P8), referring to guidelines or a company playbook (P4, P6, P9, P11), organizing documents into descriptive folders with tags (P10--P12), or using a spreadsheet or document to capture key details while reviewing (P2, P10--P12).
    To complement users' current workflows, \system{} supports creating structured representations that emulate the organizational strategies participants described, facilitating consistent review and comparison across documents.
    
 \textbf{Knowledge workers extract information using a combination of retrieval approaches, depending on the nature of the information needed to reach their goals.}
    Throughout their work, participants sometimes needed to extract a single value accurately from a single document (e.g., a date or payment amount).
    Other times they sought longer-form answers to specific questions (e.g., ``How is payment structured?''), or needed to extract query-specific excerpts from a document (e.g., ``Find all termination conditions in this contract'').
    Many participants also reported reasoning over similar information across multiple documents (e.g., ``Which candidate has the most experience?'' or ``Which contract has the ``earliest'' termination date?'').
    Overall, we found that participants sought different types of information in their documents to meet their goals, often organizing these different information types within their structured representations.
    These findings prompted \system{}'s suite of natural language \textit{actions}, which serve to emulate these document-centered foraging strategies corresponding to users' information needs. 
    For instance, \system{} helps users easily extract specific factual information (i.e., lexical search), search for document snippets semantically relevant to a query (i.e., semantic search), and answer natural language questions over documents.
    
\textbf{Extracting information across document collections is often repetitive, time-consuming, and tedious. As a result, the process is typically incomplete or prioritized by risk.}
    Participants described workflows that mainly consisted of systematic and repetitive extraction of information from their documents, reflecting prior work that has suggested the extraction and encoding of information are often the most time-consuming processes in sensemaking~\cite{russell_cost_1993}.
    Participants worked with between tens and hundreds of documents at a time, each varying in length from a few pages (<7) to hundreds (>300) and taking anywhere from a few minutes to several hours to review.
    None of the participants reported using AI assistance (e.g., LLM-powered applications) to support their document-centric tasks.
    They instead relied on established strategies, such as keyword search (i.e., Control+F) or skimming documents to manually search for the necessary information.
    However, due the length and density of jargon within business documents, these strategies were cognitively demanding and potentially haphazard, as it felt ``\textit{easy to miss a needle in the haystack}'' (P9).
    Moreover, current tools only allowed participants to search over a single document at a time, regardless of whether they needed to eventually execute similar searches over every document in their collection.
    As a result, participants reported struggling with the sheer volume of documents they encountered, often unable to review everything.
    Some employed a heuristic strategy, choosing to review only the riskiest documents, for instance by prioritizing contracts with the highest contract values (P7--P9).
    Assistance in identifying relevant information within their documents could significantly improve productivity, as one participant described: \textit{``If those [data] are made available to me, then I know where to look for. That'll reduce my time by 30-40\%, even 50\%. Because then I don’t have to read through the entire contract''} (P9).
    These findings suggest the potential for automation to both reduce tedium and improve the coverage of documents reviewed in users' current workflows.

\input{figures/cells-overview}

\textbf{While optimistic for AI assistance, knowledge workers desire agency and establish trust through manual verification.}
    AI-powered systems can provide invaluable support in information foraging, but they also inevitably err~\cite{horvitz_principles_1999, amershi_guidelines_2019} and can lack the specific expertise required for users' specialized workflows and goals.
    Participants described tasks for which they were skeptical that AI assistance could complete independently, such as those requiring complex reasoning or understanding of subtle nuances within documents.
    Participants emphasized their years of experience and contextual understanding (e.g., within a specific organization) were important in making sense of the information.
    For instance, P7 commented: \textit{``Our review from an accounting standpoint is very subjective. So it's not black and white all the time. Just because termination for convenience was found in a clause, it goes two paths---it resulted in a journal entry or it did not. Still, it's not always a journal entry.''}
    To build trust in AI assistance, participants articulated their need to understand how and why the AI arrived at a particular result.
    Participants therefore desired the ability to retain agency and saw the AI as a helpful co-pilot. They stressed the importance of having means to efficiently inspect and assess the output of AI assistance, such as through references to the \textit{``the exact language''} from a document (P5).
    \system{} addresses this need by indicating the provenance of extracted information and allowing users to edit or remove any erroneous AI responses.

\subsection{Design Goals} \label{sec:design_goals}
Summarizing our findings, we suggest an effective system for streamlining user workflows with business document collections should support the following goals:
\begin{enumerate}[label=\textbf{[D\arabic*]}]
    \item Integrate useful structured representations to help users review and organize information extracted from their documents, with efficient navigation between representations.
    \item Provide unified support for common information extraction approaches used in business document-centric workflows---lexical and semantic search, intra-document querying, and cross-document synthesis.
    \item Reduce manual efforts of repeated information foraging over dense text documents.
    \item Indicate the provenance of information, support efficient verification of reliability, and allow user control to foster trustworthy collaboration between users and AI assistance.
\end{enumerate}

%% file: tables/formative_study_participants.tex
\begin{table*}[t!]
  \small
  \centering
  \Description[Table listing participants, functional areas, and documents.]{Table listing participants, functional areas, and documents from the formative study. Functional areas of participants varied across Finance, Legal, and Administration. Documents varied across reports and contracts.}
  \caption{Participants in our formative study.}
  \begin{tabular}{p{0.04\textwidth}p{0.25\textwidth}p{0.35\textwidth}}
    \toprule
        & \textbf{Functional Area}      & \textbf{Documents} \\
    \midrule
    P1  & Financial Planning            & Company reports \\
    P2  & Finance Revenue               & Agreements \& contracts \\
    P3  & Accounting                    & Tax documents \\
    P4  & Finance Revenue               & Agreements \& contracts  \\
    P5  & Legal                         & Agreements \& contracts \\
    P6  & Finance Revenue               & Agreements \& contracts \\
    P7  & Finance Revenue               & Agreements \& contracts \\
    P8  & Financial Planning            & Agreements \& contracts \\
    P9  & Procurement                   & Agreements \& contracts \\
    P10 & Administration \& Management  & Financial reports \\
    P11 & Administration \& Management  & Classroom reports \& curriculum \\
    P12 & Administration \& Management  & Scientific reports \& information sheets \\
    \bottomrule
  \end{tabular}
  \label{tab:formative-study-overview}
\end{table*}

%% file: figures/cells-overview.tex
\begin{figure*}[t]
    \centering
    \includegraphics[width=0.95\textwidth]{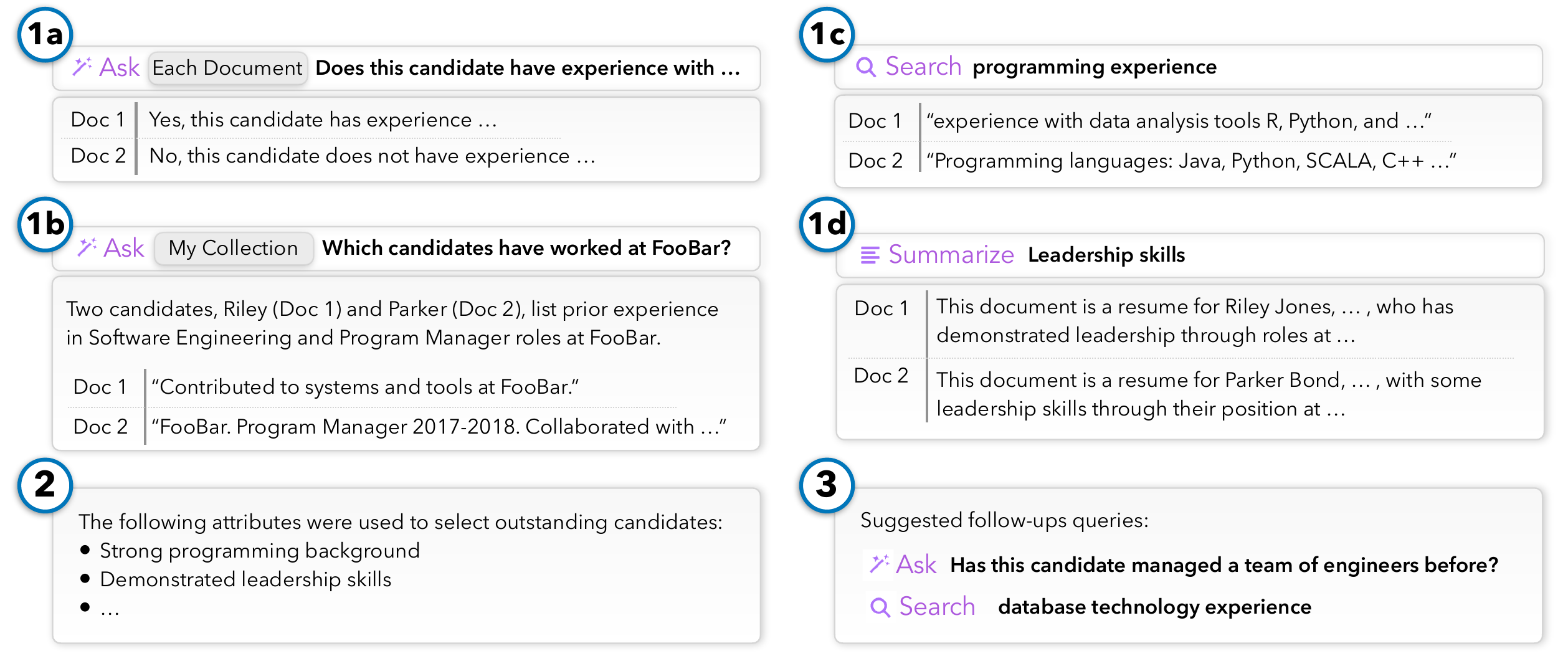}
    \caption{\system{}'s Notebook View is comprised of various cells. \textit{Action cells} provide collection-centric AI assistance for users' dynamic information needs. \texttt{Ask[Each Document]} answers the same question for each document separately (1a), \texttt{Ask[My Collection]} answers questions synthesizing information across a collection (1b), \texttt{Search} extracts information verbatim from each document (1c), and \texttt{Summarize} generates a user-guided summary for each document (1d). \textit{Text cells} serve as a note-taking space (2), and \textit{AI Suggestion cells} provide follow-up actions to continue the foraging process (3).}
    \Description [Six boxes representing cells of Marco]{Six cells used in the notebook in Marco. Four cells show different text in purple indicating its purpose: Ask Each Document, Ask My Collection, Search, and Summarize. One cell shows an example of text in an editable Text cell. One cell shows two suggested actions in an AI Suggestion cell.}
    \label{fig:cells-overview}
\end{figure*}

%% file: sections/04_system.tex
\section{The \system{} System}
Guided by the insights from our formative study, we developed \system{}, an interactive workspace that supports sensemaking by leveraging AI assistance to reduce the costs of foraging over document collections. \system{}'s user interface consists of three integrated views (Figure~\ref{fig:ui-overview}).
The Notebook View allows users to use natural language to extract information across their collection using AI assistance. As users interact with the Notebook View, foraged information is automatically aggregated in the Table View to facilitate comparison across the collection. The Document View allows users to view individual documents in their original format, providing on-demand context for extracted information. Next, we describe the design of \system{}'s user interface and interactive features, referencing our prior design goals D1--D4.

\input{figures/extract_generate}

\subsection{Notebook View}
We refer to the primary area in which users extract and schematize information with \system{} as a sensemaking \textit{notebook} (akin to the evidence file from sensemaking theory~\cite{pirolli_sensemaking_2005}). A notebook is comprised of \textit{cells}, adapting a block-based document metaphor from computational notebooks (e.g., Jupyter Notebooks~\cite{jupyter}). \system{} provides lightweight affordances for cell manipulation, allowing users to create, delete, hide, duplicate, or clear cells in a notebook. A cell can be one of three types: Text, Action, and AI Suggestion. Users create Action and Text cells on-demand, and AI Suggestion cells are suggested in response to users' actions. Figure~\ref{fig:cells-overview} highlights the different types of cells available in \system{}.

\subsubsection{Text Cell}
Text cells render rich text editors, with common tools such as lists, text emphasis, and Markdown styling. Text cells allow users to record their ongoing sensemaking processes, providing context and structure to surrounding cells. For instance, users can create in-situ notes as they explore their documents, reducing context switching to external note-taking applications.

\subsubsection{Action Cell}
To facilitate the exploration of a large document collection, \system{} equips users with AI-powered document assistance through Action cells, which execute user information-seeking queries as \textit{actions} (D3). Users create Action cells with a slash command, i.e., typing the slash character ('/'), in any empty Text cell. Three available actions are presented in a drop-down menu: \texttt{Search}, \texttt{Ask}, and \texttt{Summarize}. These specific actions are motivated by common document-centered information needs revealed in the formative interviews (D2).

\vspace{0.2em}
\textbf{\texttt{Search} Action.} The first action, \texttt{Search}, allows users to extract information from their documents relevant to one or more search queries. Results are returned as a table, with rows for documents and columns for each search query. Users can select from two types of searches: lexical and semantic. Lexical search returns short document snippets containing an exact keyword match to a user's query, while semantic search extracts short document snippets which are semantically similar to a user's query. \texttt{Search} performs semantic search by default, and users can execute a lexical search by enclosing their query in quotes. Users can also search for multiple information needs in a single \texttt{Search} action with comma-separated queries. Columns in the result table then correspond to each of the separate queries. Importantly, results returned by \texttt{Search} are verbatim snippets from each document. This both allows users to trust the provenance of the extracted information and provides a glimpse into the actual language used in the document.

\vspace{0.2em}
\textbf{\texttt{Ask} Action.} The second action, \texttt{Ask}, allows users retrieve answers to document-centered information-seeking questions. Questions are specified via natural language and \system{} returns answers based on information either within individual documents (\texttt{Ask[Each Document]}) or across all documents in the collection (\texttt{Ask[My Collection]}). Users can select from these two types of \texttt{Ask} actions depending on the context for which their question is most appropriate. The first type, \texttt{Ask[Each Document]}, allows users to ask the same question to each document in the collection, and retrieve an independent answer for each. \texttt{Ask[Each Document]} supports tasks for which users need to extract similar information across all of their documents. For instance, a hiring manager may ask their collection of resumes, ``What programming languages has this candidate used in the past?'' and expect a separate answer for each of their candidates. Responses to \texttt{Ask[Each Document]} are returned as a table, with rows for documents and a column for the query. Unlike \texttt{Search}, \system{} returns an LLM-generated answer for each document rather than an extracted snippet.

On the other hand, more complex information needs may combine both information extraction and synthesis over multiple documents. For instance, a hiring manager could ask, ``Which of these candidates have prior experience training machine learning models?'' For these types of questions, the \texttt{Ask[My Collection]} action is better suited. This action first identifies one or more pieces of information required to answer the question (e.g., ``experience training machine learning models''), extracts the relevant information from each document (using \texttt{Search}), and then synthesizes a concise answer to the question informed by the extracted snippets. The output of \texttt{Ask[My Collection]} consists of two components: first, the synthesized answer, and second, a table with the information \system{} extracted to inform its synthesized answer (same verbatim snippets as returned from \texttt{Search}), which can serve as evidence for a user to verify the system's answer or reasoning process.

\vspace{0.2em}
\textbf{\texttt{Summarize} Action.} The third action, \texttt{Summarize}, provides users with a short summary of each of their documents. As with the first two actions, results from a \texttt{Summarize} action are presented as a table with rows representing documents and a single column containing a document summary. By default, \system{} returns a general summary for each document. Users can further specify dimensions to focus on within the generated summaries. For instance, a hiring manager may want a summary of each of their candidates, but with a specific focus on their leadership skills.

Altogether, these actions meet users' different information needs, offering reliable evidence extracted from each document and generated answers to expedite synthesis in sensemaking (Figure~\ref{fig:extract-generate}).
By default, actions execute over an entire collection. However, in working toward their goals, users often begin to filter down the set of relevant information (i.e., documents) they care about. To help focus their exploration, \system{} supports drilling down into a collection by selecting a specific subset of documents to execute an action over (D2), thus returning fewer results within the table that users need to review. Finally, users can edit or remove any text within an Action cell, providing full control over inaccurate or irrelevant AI-generated results within the notebook (D4).

\subsubsection{AI Suggestion Cell}
To assist in the sensemaking process, \system{} suggests periodic guidance to users through AI Suggestion cells which contain recommendations for up to three additional information-seeking queries relevant to a users' goals (D3). To bootstrap the information foraging process, an AI Suggestion cell greets users with several starting queries when a notebook is initially created. As users accumulate action cells in their notebook, \system{} leverages their foraging history to provide relevant follow-up queries that may inspire subsequent sensemaking directions. AI Suggestion cells are non-intrusively placed below the most recently created cell, and can be accepted (i.e., turned into a pre-populated Action cell) or dismissed with a single click.

\subsection{Table View}
Individual result tables within each Action cell are automatically aggregated into an overview table within a Table View, which aims to mirror common organizational artifacts (e.g., spreadsheets) created in current business workflows.
Each row in the overview table represents one document in the collection, and each column represents one of the dimensions for which an action was created and executed in the notebook.
Using this view, users can easily compare multiple dimensions across multiple documents, reviewing their information foraging history in a single structured representation (D1). Users can also filter and reorder columns to control the exact presentation of information, and export this view to a CSV file to save, reuse, and share their work.

\subsection{Document View}
To help build trust in AI assistance, \system{} provides \textit{context linking}, an interactive feature enabling users to click on any document-grounded snippet within the other two views to open the corresponding document with attribution highlighted (D1, D4). This interaction allows extracted information to serve as an ``index'' or entry point into a document, reducing the cognitive costs of switching between document snippets in the other two views and the original source documents. Documents opened via this interaction are rendered in a Document View, placed adjacent to the other two views, and equipped with standard document functionalities (e.g, highlighting, text annotation, and keyword search).

\subsection{System Architecture}

To enable its suite of interactive features, \system{}'s architecture combines a preprocessing pipeline for ingesting collections of documents and various NLP services for executing user-created actions and suggesting follow-up actions.

\input{figures/collection-qa}

\subsubsection{Preprocessing a Collection of PDF Documents} When a collection of PDF documents is uploaded to \system{}, each document is preprocessed to minimize latency during subsequent user interactions. Documents are processed with the PDF Extract API~\cite{pdfExtractAPI}, extracting content and structural information into a structured JSON format. Sentences are split from the raw text of each document, embedded into a 384-dimensional dense representation using \texttt{multi-qa-MiniLM-L6-cos-v1} (an encoder model tuned for semantic search and trained with self-supervised contrastive learning) from the SentenceTransformers framework~\cite{reimers-2019-sentence-bert}, and indexed along with relevant text metadata using OpenSearch~\cite{opensearch}, an open-source vector database.

\subsubsection{Handling Actions over Individual Documents}
Several actions operate over an individual document (\texttt{Search}, \texttt{Ask[Each Document]}, \texttt{Summarize}), but are repeated for each document in the collection independently. For each document, we first retrieve relevant context (i.e., 30 chunks with the greatest cosine similarity to the embedded query). The retrieved chunks are sorted by the order they appear in the document and then concatenated with a query to form a few-shot prompt for an LLM (\texttt{gpt-3.5-turbo}). The specific formatting of the query depends on the action type. For instance, \texttt{Ask} uses a user's query verbatim, while \texttt{Search} uses a template, ``Search for \{user's query\}.'' To allow for a more responsive user experience, we minimize latency by parallelizing across documents and streaming back LLM responses for each document independently. In this way, users can typically start reviewing information from their documents within 1--2 seconds. We opt to use the \texttt{gpt-3.5-turbo} model for these actions (instead of more performant models, e.g., \texttt{gpt-4}) due to its lower cost and latency.

\subsubsection{Answering Queries with Collection Context}
To handle actions that need to synthesize information across a collection (e.g., \texttt{Ask[Collection]}), we design a multi-phase prompting approach (Figure~\ref{fig:collection-qa}), inspired by similar decomposition strategies for answering complex queries with LLMs (e.g.,~\cite{radhakrishnan2023question, khot2023decomposed}). Based on a user's query, \texttt{gpt-4} is first prompted to identify the relevant document attributes necessary to answer the query. These attributes could include some a user had already previously searched for, in addition to other missing attributes. For each missing attribute, \system{} executes a \texttt{Search} action over each individual document, extracting and saving relevant information into an evidence table. The context for a second prompt is then formed by joining the relevant information from each document for each of the identified attributes. Finally, \texttt{gpt-4} is prompted using the context and original query.

\subsubsection{Generating Suggested Actions}
AI suggestions for initial or followup actions are generated using a few-shot prompting strategy with an LLM (\texttt{gpt-3.5-turbo}). The prompt incorporates the user's goal, up to three documents from the document collection (truncated to the first thousand characters of full text), and the user's interaction history (i.e., any prior queries to an action cell).

\subsubsection{Implementation}
\system{} was built as a standalone web application, with JavaScript and React~\cite{react} for user interface components. All preprocessing and language understanding services were implemented in Python, using zero-shot and few-shot prompting with LLMs accessed through OpenAI APIs. LLMs were prompted with a sampling temperature of 0 and max generation length of 256 tokens (except AI Suggestion cells, for which a temperature of 0.7 and max generation length of 128 tokens were used). Exact prompts for all actions are provided in Appendix~\ref{appendix:llm-prompts}.

%% file: figures/extract_generate.tex
\begin{figure*}[t]
    \centering
    \includegraphics[width=0.9\textwidth]{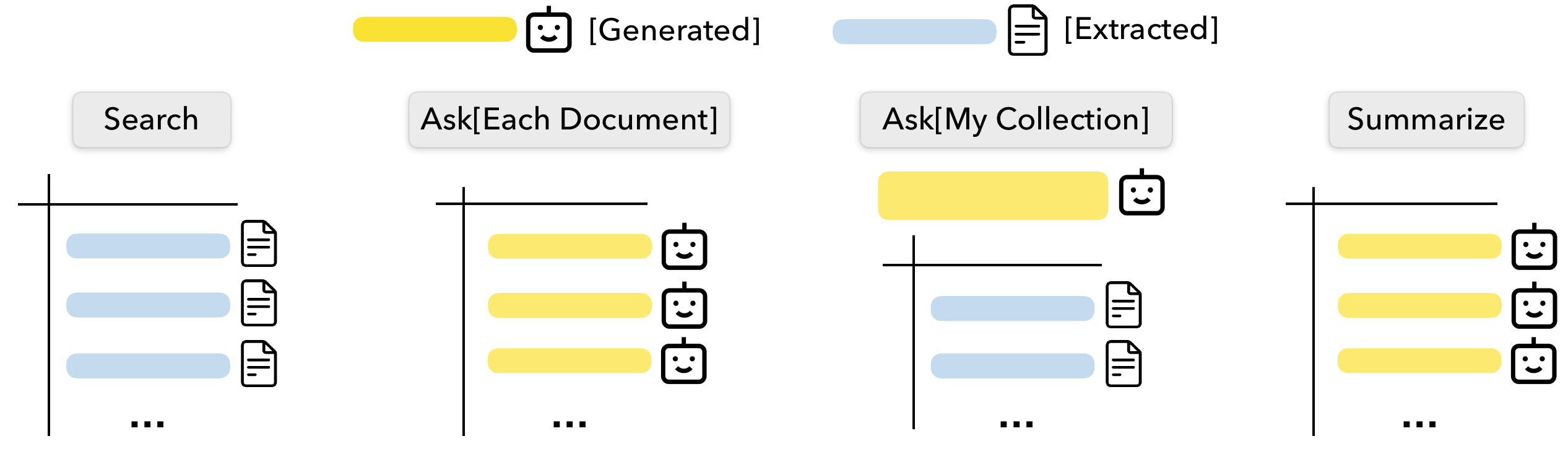}
    \caption{\system{} supports four strategies for information foraging across a collection. \texttt{Search} returns snippets extracted directly from each document, \texttt{Ask[Each Document]} and \texttt{Summarize} return LLM-generated answers grounded in each document, and \texttt{Ask[Each Document]} returns a combination of both extracted evidence and an LLM-generated answer synthesizing the evidence.}
    \Description[Representation of extracted and generated content in Action cells]{There are representations of four actions cells. From left to right: search, with blue boxes indicating extracted content; ask each document, with yellow boxes indicating generated content; ask my collection, with both yellow box indicating generated answer and blue boxes indicating extracted content; summarize, with yellow boxes indicating generated content.}
    \label{fig:extract-generate}
\end{figure*}

%% file: figures/collection-qa.tex
\begin{figure*}[t]
    \centering
    \includegraphics[width=0.95\textwidth]{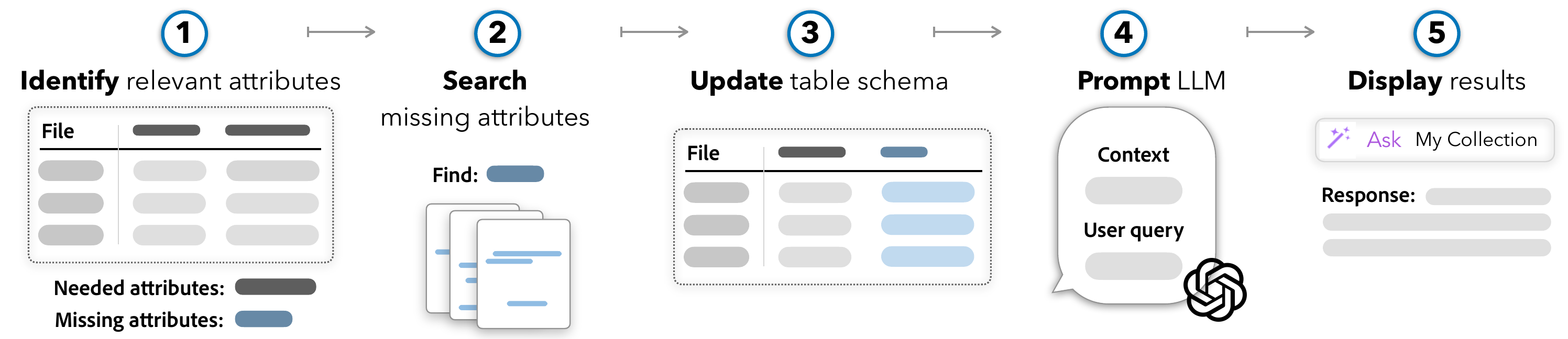}
    \caption{Overview of \system{}'s multi-step pipeline for answering users' queries over a document collection. \system{} first identifies attributes required to answer the query, some of which may already have been retrieved by prior user queries and other which may be missing (1). For each missing attribute, \system{} executes a search to extract relevant snippets of information from each document (2), and saves the new search results into the aggregate table (3). Search results for each of the required attributes are formatted into a prompt and sent to an LLM (4), whose response to displayed to the user (5).}
    \Description[Flow diagram with five steps]{A flow diagram with five steps. The steps numbered 1 to 5 and connected by arrows, from left to right are: Identify relevant attributes, Search missing attributes, Update table schema, Prompt LLM, and Display Results. A visual representation of each action is shown below the text.}
    \label{fig:collection-qa}
\end{figure*}

%% file: sections/05_evaluation.tex
\section{Study 1: Controlled Usability Study} \label{sec:usability-study}

To evaluate the efficacy of \system{}'s design and interactive features (RQ1 and RQ2), we conducted a controlled usability study, comparing \system{} and a baseline approach. We used a $2 \times 1$ within-subjects study design, counterbalancing system conditions across participants to control for order effects and minimize biases. Participants completed one task with \system{} (\textsc{Marco}) and one task with the system default file manager and document viewer (\textsc{Baseline}). The baseline condition was chosen to reflect the current workflows participants described in our formative study, namely the use of a text editor or spreadsheet (e.g., Google Docs and Sheets) and a PDF reader (e.g., Acrobat Reader or system default).

\subsection{Participants}
We recruited 16 participants (9 female, 7 male; average age of 28.2, SD = 4.5) via group messaging channels within a large software organization. Participants consisted of 6 software engineers, 3 PhD students, 2 business analysts, 1 Master's student, 1 program manager, 1 UX designer, 1 learning architect, and 1 sales employee. Participants were required to be 18 years or older and able to read documents written in English. All participants reported having previously used some LLM-powered application (e.g., ChatGPT), and three participants indicated experience developing applications or conducting research with LLMs.

\subsection{Procedure} \label{sec:usability-study-procedure}
We first introduced the study and obtained consent. We then guided participants through an interactive tutorial of \system{}, highlighting the intended usage and limitations of \system{}'s AI-powered features. Afterwards, participants were asked to role-play as knowledge workers in two scenarios with business document collections, completing the following two tasks:
\begin{enumerate}
    \item \textit{Hiring}. Participants assumed the role of a hiring manager for an entry-level technology analyst position at a financial company, and were given a collection of 15 candidates' resumes. Resumes were one to two pages long and curated from publicly available university resume books.
    \item \textit{Cleaning}. Participants assumed the role of an office manager for a start-up company looking for a new cleaning service provider, and were given a collection of 10 contracts from potential providers. The contracts were pulled from online templates and modified to be two to four pages long.
\end{enumerate}
To ensure both tasks could be completed in a single study session and also include a non-trivial number of documents, we adjusted the content and length of several documents. We also used fewer contracts in the \textit{Cleaning} task since contracts tended to be longer than the resumes in the \textit{Hiring} task.

For each of the two tasks, participants answered three questions. One question was intended to be more conducive to keyword search, while the other two questions involved reading and reasoning over more of each documents' text. Additional details for the two tasks can be found in Appendix \ref{appendix:usability-study}. Each question asked participants to select one or more documents in the collection as their final answer. Participants were limited to seven minutes per question, allowed to submit before time expired, and submitted partial progress if time expired. The allotted time was iteratively determined through several pilot studies. The final time limit was chosen to allow most participants to manually review each document in \textsc{Baseline} when completing the tasks. We believe reducing the allotted time would increase the difficulty of the task and thereby the comparative advantage of \system{}, but we leave these studies for future work.

After both tasks were completed, participants completed a system usability survey and a demographics survey. Any remaining time in the study was used to allow participants to share their overall experience using \system{} and provide additional feedback. All interviews were recorded, transcribed, analyzed for qualitative insights following an open thematic analysis~\cite{braun2006using, boyatzis1998transforming}. Studies ranged between 60 and 70 minutes long. Participants were compensated with \$30 (USD) upon completion of the study.

\subsection{Measures}
We recorded the following measures for each question:
\begin{itemize}
    \item \textsc{Accuracy} -- A score from 0 to 1 indicating overlap between a participant's final answer and a predetermined ground truth answer. \textsc{Accuracy} was calculated as the sum of true positives and true negatives, divided by the number of documents in the associated collection for the question.
    \item \textsc{Time} -- The amount of time a participant spends answering a question, measured from when a participant finishes reading a question to submission of a final answer.
    \item \textsc{Confidence} -- A participant's confidence in their answer. Specifically, their response to the question ``How confident do you feel in your answer?'' on a 5-point Likert scale ranging from ``Not at all confident'' to ``Extremely confident.''
    \item \textsc{Ease} -- A participant's ease of identifying the information used to arrive at their answer. Specifically, their response to the question ``How difficult was it to find the information you needed to reach an answer?'' on a 5-point Likert scale ranging from ``Extremely easy'' to ``Extremely difficult.''
    \item \textsc{Effort} -- A participant's amount of cognitive effort required to arrive at their answer. Specifically, their response to the question ``How mentally demanding was completing the task?'' on a 5-point Likert scale ranging from ``Extremely low'' to ``Extremely high.''
\end{itemize}
We also measured overall usability with a system usability scale consisting of ten 5-point Likert scale questions~\cite{brooke1996sus}. 

\subsection{Analysis}
We analyzed \textsc{Time}, \textsc{Confidence}, \textsc{Ease}, and \textsc{Effort} using linear mixed-effects models (LMMs)~\cite{raudenbush2002hierarchical}, which allow both fixed and random effects (e.g., individual participants) and hierarchical data (e.g., non-independent participant observations). For \textsc{Accuracy} with response values ranging from 0 to 1, we used a binomial generalized linear mixed-effects model (binomial GLMMs). Both LMMs and GLMMs are commonly used to analyze similar measurements from studies across medicine~\cite{cnaan1997using}, behavioral science~\cite{cudeck1996mixed}, and human-computer interaction~\cite{do_err_2023,siu_supporting_2022}. We used likelihood ratio tests of a full model with the effect in question against a reduced model without the effect in question to obtain p-values, a standard method for significance testing with LMMs~\cite{vuong1989likelihood}. Models were fit using the \textsc{lme4} package in R~\cite{bates_fitting_2015}.

\subsection{Results}

\subsubsection{Efficacy of \system{} as Sensemaking Support (RQ1)}
Figures~\ref{fig:results-quant} and~\ref{fig:results-likert} summarize how quantitative measures for the task questions varied between the two system conditions. A main effect analysis of system condition on \textsc{Time} found participants completed questions more quickly with \textsc{Marco} than with \textsc{Baseline} ($M = 297.3s$, $SD = 89.1$) ($\chi^2(1) = 8.26, p = .004$). On average, participants completed each question with \textsc{Marco} in 249.6s (SD = 88.8s) and with \textsc{Baseline} in 297.3s (SD = 89.1s), a 16\% difference. Participants were slightly more accurate with \textsc{Baseline} (91.9\% accuracy, SD = 8.3) than with \textsc{Marco} (88.1\% accuracy, SD = 10.1), though this difference was not significant ($\chi^2(1) = 3.35, p = .067$). A similar analysis for system condition on \textsc{Confidence} found no difference ($\chi^2(1) = 1.31, p = .25$) in participants' self-reported confidence in their answers between \textsc{Marco} (Mdn = 3, IQR = 3-4) and \textsc{Baseline} (Mdn = 4, IQR = 3-4). A main effect analysis of system condition on participants' self-reported measures for both \textsc{Ease} ($\chi^2(1) = 27.73, p < .001$) and \textsc{Effort} ($\chi^2(1) = 33.97, p < .001$) showed participants found it easier to locate information necessary to answer the task questions with \textsc{Marco} (Mdn = 2, IQR = 1-3) than with \textsc{Baseline} (Mdn = 3, IQR = 3-4), and found task questions required less mental effort to complete with \textsc{Marco} (Mdn = 2, IQR = 1-3) than with \textsc{Baseline} (Mdn = 3, IQR = 2-4).

\input{figures/results_quant}
\input{figures/results_likert}

\subsubsection{Participants' Usage Patterns of \system{} (RQ2)}
We provide insight into usage patterns with \system{} through an analysis of participants' interaction logs and semi-structured interview responses. We refer to participants with the pseudonyms P1–16. Interaction counts are provided in Appendix Table~\ref{tab:interaction-counts}. In the \textsc{Baseline} condition, participants completed tasks by opening and scanning each document in turn, often relying on keyword search and structural and visual cues (e.g., section headings) to identify passages to read. Participants using \system{} tended to have fewer interactions with individual documents. They instead created \texttt{Search} and \texttt{Ask} actions, the specific type and quantity of which varied across participants. On average, across three questions, participants created 2.3 (SD = 1.3) \texttt{Search}, 2.4 (SD = 1.6) \texttt{Ask[Each Document]}, and 1.7 (SD = 1.3) \texttt{Ask[My Collection]} actions. Participants preferred \texttt{Search} when comparing multiple criteria across their collection at once, searching for 2.0 queries per action on average (SD = 0.94, Mdn = 2.0), and up to as many as six queries in a single action (P16).

\subsubsection{System Usability and Suggested Improvements}
The average and median SUS scores were 74.5 and 71.3 respectively, indicating strong overall usability.
For instance, some participants found the need to select an action appropriate to a specific information need added cognitive overhead. Instead, based on a user's query, the system could determine which type of action cell to create and whether generative or extractive results is more likely desired (P2--P5). Others suggested providing clearer visual indicators to distinguish between LLM-generated text from \texttt{Ask} and extractive document snippets from \texttt{Search} (P1, P14, P15).

\section{Study 2: Design Probe with Knowledge Workers}

We conducted a second qualitative study using \system{} as a design probe with domain experts whose responsibilities included reviewing large sets of documents. The goal of this study was to understand how \system{} might support their current real-world workflows (RQ2) and how they perceive and interact with potentially imprecise AI assistance (RQ3).

\subsection{Participants}
We recruited 7 participants (3 female, 4 male) via group messaging channels within a large software organization (Table~\ref{tab:design-probe-overview}). Unlike in the usability study, we required participants' work responsibilities to include reviewing large sets of documents. Two of the participants had between 1--5 years of experience, and the remaining five participants had more than 15 years of experience working with business documents. One participant reported using LLM-powered applications to support his work (P1). Participants were thanked for their time but not compensated.

\subsection{Procedure and Analysis}
The study began with introductions and a discussion of the participants’ roles, responsibilities, and goals when reviewing document collections. Next, a study facilitator guided participants through a walkthrough of \system{} using a document collection of 10 contracts from the usability study (\S\ref{sec:usability-study-procedure}), highlighting \system{}'s features. At each step, participants were encouraged to talk through what they encountered and their reactions. The study concluded with a semi-structured interview to understand how \system{} could support participants’ current workflows they described at the start of the study. For confidentiality reasons, we opted to demonstrate \system{}'s capabilities with a pre-selected document collection instead of users' own work documents. Interviews were conducted remotely, recorded, and transcribed for analysis. One author went through the transcripts and coded them for themes using an open thematic analysis process~\cite{braun2006using, boyatzis1998transforming}. The research team then discussed and iterated upon the themes until consensus.

\subsection{Results}

\input{tables/design_probe_overview_v2}

Participants reiterated similar information needs and challenges in their current workflows as those found in our formative study (\S\ref{sec:formative-study-findings}).
For instance, one operations analyst described how his team often receives 15--20 documents every 30 minutes to review within 24 hours (P2). To meet their goals, all participants described spending considerable time and manual effort extracting information from various documents.
Given the substantial volume and cognitive load, participants prioritized their focus, saying, \textit{``We don't check everything. That's impossible.''} (P6). Next, we report how \system{} could support participants' current workflows and identify areas for improvement. We organize our findings under themes that emerged from our qualitative analysis.

\subsubsection{Accelerating information foraging and facilitating re-use}
Participants described having a basic set of recurrent questions they needed to answer over multiple documents and analyses (P1--P4, P6, P7): \textit{``We just go through our checklist of five or six questions and then mentally just go through the document and try to answer it''} (P2). \system{} could lower the foraging costs for repetitive questions, enabling participants to instead focus on validating and analyzing the results. Participants also appreciated how \system{} provided structure to reuse analyses (through defined actions) over their documents and apply the same analyses over a new set of documents.

\subsubsection{Different actions support different sensemaking use cases}
Most participants appreciated having different types of actions to collect information, which supported various use cases (P1, P2, P4, P5). In some cases, a generated answer over the collection (\texttt{Ask[My Collection]}) could provide participants with sufficient detail, while for other analyses, a \texttt{Search} action with document snippets better supported users’ goals of familiarizing with each document's language. A finance analyst described how \system{}'s design aligned well with existing workflows, for instance with the two types of analyses performed by her team---recommending a strategy and cross-checking details---\textit{``I love the fact that it's all documents, each document, a custom set of documents that you want where you could choose \dots that's really important because that's how people are working''} (P4). One participant, a legal specialist, found the distinction between actions unnecessary, and suggested \system{} could instead understand users’ intents based on their queries (P3).

Participants had varying preferences for which actions would best support their unique workflows. Those whose workflows involved processing document-by-document found most use for \texttt{Ask[Each Document]} since it closely matched their working mental model (P1, P2, P7). \texttt{Ask[My Collection]} was seen as useful for making comparisons (e.g., different contract vendors) (P1, P6), identifying documents with relevant terms or statistics (P2, P4), and uncovering patterns across documents (P5). Finally, \texttt{Search} provided an additional mechanism for confirming results (P2, P3, P5). Table~\ref{tab:design-probe-overview} lists some of the specific actions participants desired.

\subsubsection{Domain experts preferred Notebook View for analysis and Table View for synthesis}
By extracting and aggregating information in one organized workspace, all participants described \system{} was better at supporting sensemaking tasks (e.g., drawing comparisons, observing patterns, and identifying non-standard language) compared to their current workflows. Participants described the Notebook View as better suited for deeper analysis and exploration of actions and queries to use with different types of analyses (P3, P4). On the other hand, the Table View provided a better overview of the information once the analyses were complete (P3, P4, P7) and subsequently was seen as more useful for \textit{``day-to-day operational tasks''} (P3). A global sourcing analyst suggested the Table View columns could be fixed and queries re-applied for day-to-day diagnostics, \textit{``If you have a few questions, most important things you identify, ask those questions and then you just change it to this view and you can export this showing the most relevant information side by side. I like that.''} (P1). To improve its utility, participants suggested being able to save queries as presets in the Table View and adding customizable column filters for different types of analyses (P3, P4).

\subsubsection{Verifiable AI assistance can provide value despite imperfections}
Participants described how \system{} could help augment their current processes, even if the AI assistance was imperfect (P2, P4, P6, P7). \textit{``Even if it's not 100\%, even just catching stuff we might miss''} (P2). For most users using \system{} was described as just another round of review (out of several) and not a replacement to their final review. An operations manager described it as, \textit{``I feel like this AI chatbot is more to help us \dots The actual work is being done by us''} (P7). She elaborated that in real use, when an analyst cannot find some information, they would raise a concern for additional review to a colleague or manager. From this perspective, \system{} was seen as another reviewer in their operations.

Document highlights that connected to results provided by \system{} were important in building confidence when using the system. Highlights made verification easy when needed. Participants emphasized that for risky business cases, verification would always be required (P4, P6). Highlights were also seen as helpful to guide further reading when generated answers lacked detail, \textit{``It’s useful because maybe the answer we get might be small, just a summary \dots Linking back will give us a step-by-step, a detailed process''} (P7).

Participants suggested ways \system{} could increase their confidence in its generated responses. First, the system should ensure responses are consistent for similar queries from different actions. Participants might apply multiple similar queries to arrive at the same answers. Providing several mechanisms to verify responses increases users’ confidence (P2, P3, P5). Second, the system should offer lengthier answers that provide explanations, which participants preferred, over succinct answers. Knowledge work involves understanding not just reporting a correct answer. As one participant described, \textit{``I want the AI to help me actually understand it and not just provide me the answer''} (P1). Third, the system should avoid overconfidence and communicate when human review is needed (P1, P2). Participants viewed \system{} as an additional thought partner in their workflow, and akin to working with other colleagues perfection was not expected so it should ask for help when needed. For instance, they suggested \system{} could communicate uncertainty in answers or flag items requiring further human review.

%% file: figures/results_quant.tex
\begin{figure*}[t]
    \centering
    \includegraphics[width=0.9\textwidth]{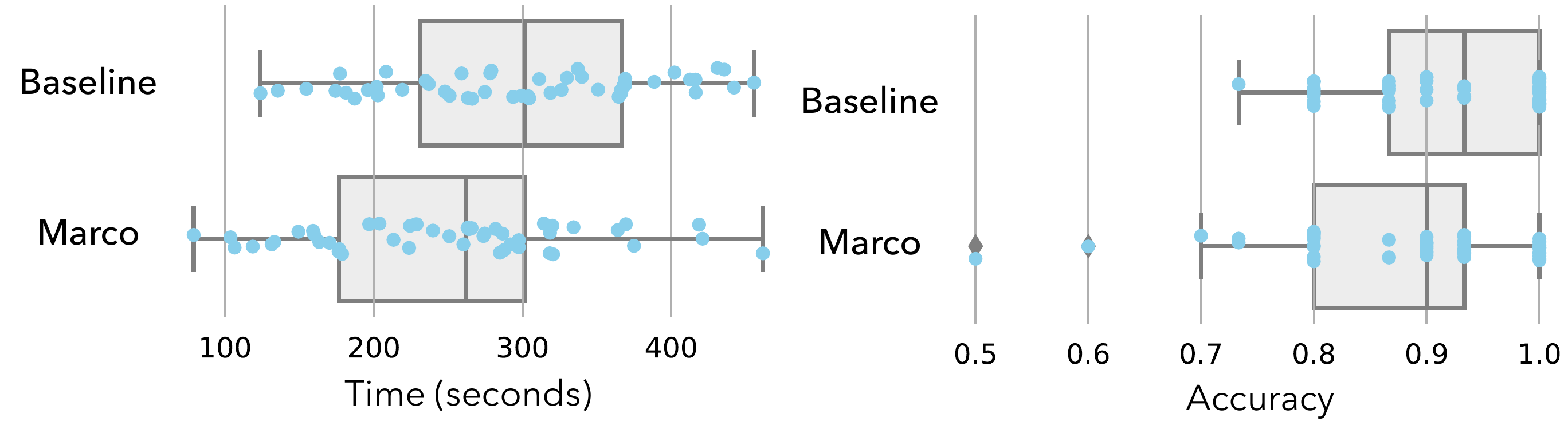}
    \caption{Quantitative results for task questions. Participants completed tasks more quickly ($p < .01$) with \textsc{Marco} ($M = 249.6s$, $SD = 88.8s$) than with \textsc{Baseline} ($M = 297.3s$, $SD = 89.1s$). Participants were slightly more accurate with \textsc{Baseline} ($M=0.92$, $SD=0.08$) than with \textsc{Marco} ($M=0.88$, $SD=0.11$); this difference was not significant at $\alpha = .05$.}
    \Description[Two stacked boxplots]{Two graphs, each containing two stacked horizontal boxplots. The first graph shows Time in seconds, with a median value for Baseline at about 300 with most values between 225 and 375, and a median value for Marco at about 260 with most values between 180 and 300. The second graph shows Accuracy, with a median value for Baseline at 0.93 and a median value for Marco at 0.9.}
    \label{fig:results-quant}
\end{figure*}

%% file: figures/results_likert.tex
\begin{figure*}[t]
    \centering
    \includegraphics[width=0.97\textwidth]{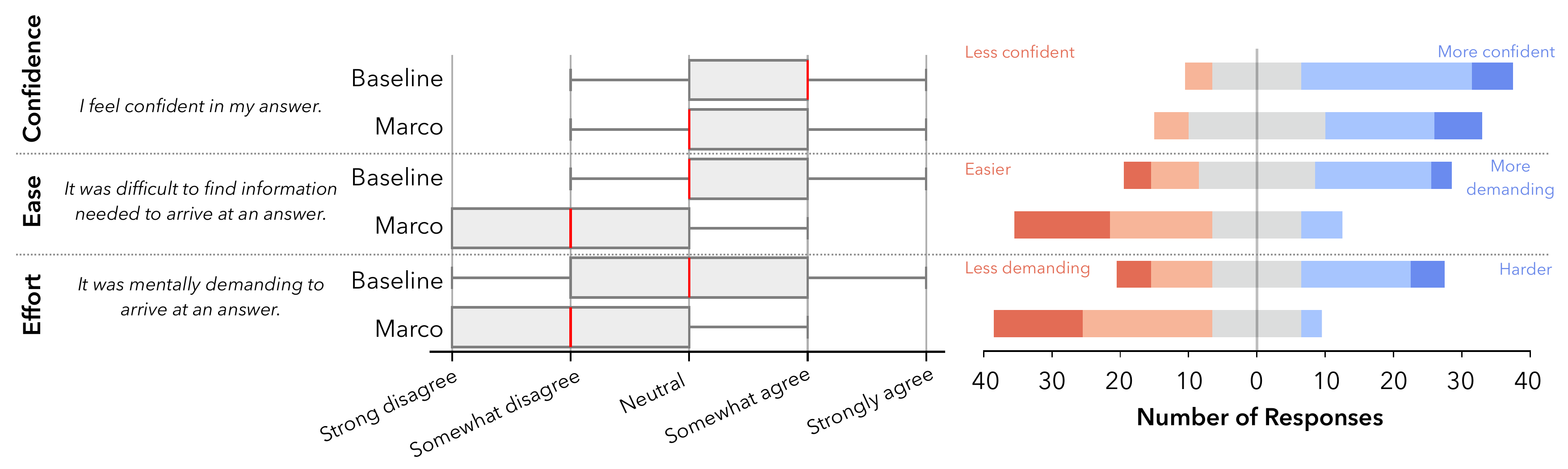}
    \caption{Subjective responses for task questions. Participants found tasks easier and less effortful with \textsc{Marco} than \textsc{Baseline}.}
    \Description[Six stacked diverging bar graphs for Likert values]{Six stacked diverging bar graphs for self-reported task measures with Likert scores. The bar graphs are separated into three groups of two, labeled with Confidence, Ease, and Effort. The two graphs for Confidence look similar, while the two graphs for Ease and Effort show more values to the left of the center line for Marco than for Baseline.}
    \label{fig:results-likert}
\end{figure*}

%% file: tables/design_probe_overview_v2.tex
\renewcommand{\arraystretch}{1.3}
\begin{table*}[ht]
  \small
  \centering
  \Description[Table listing participants, primary document collections, and desired actions with Marco.]{Table listing participants, primary document collections, and desired actions with Marco. Seven participants are listed. Collections ranged across contracts and documentation. Desired actions list various job-specific questions and search queries.} 
  \caption{Overview of knowledge workers in the design probe, their documents, and how they would use \system{}.}
  \begin{tabular}{p{0.02\textwidth}p{0.2\textwidth}p{0.7\textwidth}}
    \toprule
     & \textbf{Functional Area} $\vert$ \newline \textbf{Document Collections} & \multirow{2}{*}{\textbf{Desired Actions with \system{}}} \\
    \midrule
    P1 & Procurement $\vert$ Hardware contracts &
        \textbf{Ask:} How quickly do we need to pay after invoicing?
        What are obligations in terms of pricing? 
        What is my fixed capacity?
        What sort of rate protection do we have? 
        What is the term growth?
        What is the cost?
        What products are included?
        Who is the supplier?
        What risks are associated with <clauses>?
        Do these contracts have any unfair or unfavorable provisions?
        \newline
        \textbf{Search:} payment obligations \newline
        \textbf{Summarize:} deals and analyses of suppliers and vendors \\ 
    P2 & Billing Operations $\vert$ Signed contracts, purchase orders, process documents, addendums & 
        \textbf{Ask:} Which documents discuss <action>?
        Which of these deals have the <identifier number>?
        Which of these deals have a flexible start date?
        Which of these deals have a keep end date pricing?
        How many <identifiers> are listed?
        What are unit prices? 
        Does it say auto-renewal? 
        \newline
        \textbf{Search:} net payment terms, start and end dates \\ 
    P3 & Legal $\vert$ Contracts & 
        \textbf{Ask:} Which contracts have <X> days notice period for termination? What are all agreements that include a DPA? 
        Who are all customers that have analytic products and less than 30 day notice period for breach? \newline
        \textbf{Search:} <agreement condition> \\
    P4 & Finance Systems $\vert$ Earning call reports, contracts & 
        \textbf{Ask:} Which contracts have <agreement terms e.g., termination for convenience>? 
        How frequent was <term> mentioned? 
        What questions are asked on <X>? 
        Which of the <company> contracts do we need to let roll through the cycle?
        What is the commencement date? \\
    P5 & Cloud Operations $\vert$ Vendor contracts, vendor documentation, system outage reports & 
        \textbf{Ask:} 
        Is there guidance on <event outage>? 
        Have you seen <event outage>? In the context of these docs, should I do <event action>? 
        Do you have an opinion on what do when <event outage> happens? 
        Which have <X> terms? 
        Which contracts are expiring? 
        What is the opt out clause? 
        How do we pay?\newline
        \textbf{Summarize:} various contract terms \\ 
    P6 & Contract \& Business Operations $\vert$ Contracts & 
        \textbf{Ask:} What is the contract number? \newline
        \textbf{Search:} payment terms, renewal terms \\ 
    P7 & Order Management $\vert$ Process documents, contracts &
        \textbf{Ask:} Which process documents are about <negotiation deal>? 
        Where is <agreement term>?
        Which agreements have an addendum changing a billing cycle? \\
    \bottomrule
  \end{tabular}
  \label{tab:design-probe-overview}
\end{table*}

%% file: sections/06_discussion.tex
\section{Discussion}
This work was motivated by knowledge workers' need to extract and analyze information from a collection of business documents to solve complex information tasks. Our interviews with knowledge workers in various business functions describe participants' diverse document-centric workflows. Despite this diversity, we identify common challenges with information extraction that motivate the design of \system{}. Our observations highlight how current tools at most enable users to access information in documents individually, resulting in tedious and repetitive processes when attempting to complete sensemaking tasks over a collection. We reaffirm that adoption and use of document-centered automation tools in the workplace remains limited~\cite{jahanbakhsh_understanding_2022}, as we also found limited use of reading assistance beyond interfaces with simple document annotation and manipulation. A study by~\citet{adler1998diary} in the late 90s provides a detailed characterization of knowledge workers’ reading activities at a time when processes where just starting to move from paper to digital. We found that while document processes described by participants were mostly digital, challenges with information extraction persist. We hypothesized these challenges remain because most digital reading tools mirror manual \textit{paper-like} reading metaphors. Our goal with \system{} was to consider an interaction paradigm in which knowledge workers primarily interact with their documents altogether as a holistic knowledge base rather than as individual documents.

To support knowledge workers' wide range of goals, \system{} was designed such that users could communicate their information needs to an AI assistant using natural language and interactively build a schema that provides an overview of a document collection. Users remain in control over what type of information is relevant for their specialized analyses but delegate the extraction and organization of information to an AI assistant through actions. We found \system's approach had a positive impact on users' productivity. Results from the controlled usability study found participants completed information-seeking tasks 16\% faster and with less effort using \system{} when compared to a manual approach. One participant reflected on how using \system{} reduced the overall workload, saying \textit{``I definitely feel it saves a lot of my time of going back and forth in the document to search stuff, so it’s really helping me in the kind of mental demand and the amount of work I need to do.''} (P1). While the various actions in \system{} provided users flexibility in addressing diverse information needs, some users experienced a learning curve due to the distinctions between these actions and having to craft the right prompt. Future work can explore how AI assistance could help users surface their intent or refine their prompts through better conversational repair strategies~\cite{albert2018repair}.
 
Ensuring the reliability of imprecise AI assistance is a common strategy by users to mitigate the impact of errors in AI-enabled interactions~\cite{amershi_guidelines_2019, fok_search_2023}. In both the usability study and design probe with experts, we observed two common strategies participants used to build confidence with \system{} and to recover from errors. First, we observed participants often used different actions in concert for a single query (e.g., both \texttt{Search} and \texttt{Ask}) as a means to cross-check retrieved results and surface any inconsistencies. Domain experts described how in workplace contexts where complex information tasks drive high-impact decisions, multiple people would be working together to verify information in different ways, and \system{} could provide an additional redundant layer of review to inform their decision making. Second, we observed when participants had low confidence or encountered any inconsistency, most opted to manually verify the accuracy of AI responses. Action cells in \system{} provided an organized table of results with evidence from each document (e.g., extractive document snippets in \texttt{Search}). This structure allowed users to quickly skim results across all documents and link to specific documents for additional verification when needed. In the usability study, few participants completed the tasks without opening a document and we found no significant differences in task accuracy and self-reported confidence when using \system{} compared to \textsc{Baseline}. These observations imply that \system{} provided adequate mechanisms for users to build trust and reliance on delegated tasks to \system{}, despite not manually reading and reviewing each document individually.


Taken together, findings from both studies suggest that \system{} can provide users a productivity boost without impacting work quality. 
Hearkening back to the ``overview first, zoom and filter, details later'' mantra for visual information seeking~\cite{eyes_schneiderman_1996}, knowledge workers using \system's collection-centric interaction paradigm need only attend to and reason over the relevant portions of each document, delving into specific documents for details as needed. When relying on AI assistance, the ability to expand on relevant document details is especially important in supporting human evaluation and verification.
However, additional studies are needed to understand real-world scenarios where overreliance on AI results may start to impact quality. For instance, several domain experts described scenarios where a review may entail working with thousands of documents and/or tight timelines, making it impossible to manually verify all document details. The strategies we observed with \system{} for cross-checking and verification could also become ineffective in these scenarios and potentially encourage overreliance~\cite{ashktorab2021ai}. Future work can investigate additional interactions that provide guardrails and mitigate risk in these high-stakes scenarios, for example by providing explorable uncertainty visualizations or further scaffolding results through clustering~\cite{cheng2019explaining, kocielnik2019will, ashktorab2021ai, felix2018exploratory}.
Our findings moreover illustrate how AI assistance can reshape traditional foraging and sensemaking activities. In these AI-augmented sensemaking processes, humans---previously burdened with cognitively taxing and tedious foraging tasks---now bear a different set of responsibilities, of specifying intents, delegating processes, and evaluating the results of AI assistance.

\subsection{Limitations}
Most of the document-centric workflows in our studies center around information in long-form, primarily text, documents. Thus \system{} was designed to primarily support text information foraging. Document workflows can also involve multimodal content (e.g., visual, structural, auditory). For instance, an analyst designing a new marketing campaign might need to synthesize information present in text, image, or video content. We see exciting opportunities in leveraging multimodal LLMs to support interactions over more diverse content~\cite{wu2023multimodal}. These interactions could expand on the types of action cells used for information foraging, as well as the representations used to organize that information (e.g., data visualizations, graphics)~\cite{chen2022crossdata}.

In our controlled usability study, we limited the number and length of documents in each collection. Task questions were also designed to be answerable within seven minutes, ensuring completion within an hour-long session. Despite these constraints, we believe the specific document categories and task questions capture the types of information-seeking queries found in our formative study. Many workflows required extracting information from documents according to established domain-specific criteria and organizing the information into a representation to share with others. Our evaluations were therefore designed around the challenges inherent in these tasks. In contrast, other sensemaking tasks that are largely exploratory (e.g., learning a new skill, debugging code, or creative writing) may not be as ideally compatible. Nevertheless, we believe \system{} can offer some utility toward these exploratory tasks. As users iteratively brainstorm new dimensions to evaluate, \system{}'s suite of actions allows users to quickly extract and compare relevant information for each dimension across many documents.

Beyond exploratory tasks, complex information tasks where needs may not be as well-defined could span multiple hours or days. As constraints on our controlled study could have affected participants' information foraging and decision making behavior, in future work we intend to conduct a longitudinal field study with \system{} involving a diverse set of business-related knowledge workers and document collections. This in-the-wild study could offer additional insights into scenarios where \system{} enhances users' workflows and areas where it may fall short.

%% file: sections/07_conclusion.tex
\section{Conclusion}
We presented \system{}, a mixed-initiative workspace leveraging large language models to support workflows with business documents. Through a suite of natural language actions that reflect common information-seeking approaches to document review, \system{} aims to improve consistency and reduce tedium in searching for information across many documents. A usability study found \system{} helped people search for and reason over information across document collections more quickly and with less effort compared to an existing baseline approach. A design probe with knowledge workers further showed how the design of \system{}'s actions and views aligned with real-world workflows. Overall, our studies highlight how business workflows even today remain manual and tedious, with low adoption of technological support. We believe \system{} offers a glimpse into how document-centered AI assistance can be integrated to complement users' processes, with simple affordances that build appropriate trust in automation. We hope this work can inspire future exploration of the opportunities offered by recent automated language understanding capabilities and envision new mixed-initiative systems for document-centered assistance.

%% file: sections/08_appendix.tex
\appendix

\section{System Implementation Details}

\subsection{\system{}'s User Interface} \label{appendix:ui-screenshots}
Figures~\ref{fig:ui_pdf_with_notebook},~\ref{fig:ui_notebook_view}, and~\ref{fig:ui_table_view} provide screenshots from \system{} to highlight common configurations of views in the user interface.

\input{figures/ui_pdf_with_notebook}
\input{figures/ui_notebook_view}
\input{figures/ui_table_view}

\section{Usability Study Details} \label{appendix:usability-study}

\subsection{Apparatus} Each participant completed the study on a provided Apple MacBook M1 Pro with 16GB of RAM running MacOS Ventura 13.4. No external monitor was used. Before beginning the study, participants were instructed on how to quickly navigate between desktops, as the Qualtrics survey for collection of task responses, the file manager and PDFs used in the control condition, and \system{} were provided on separate desktops. All participants were comfortable with using a computer to browse documents and expressed little to no difficulty using the provided apparatus.

\subsection{Scenario-based Tasks}
Participants in the usability study completed the following two scenario-based tasks. For each task, they answered three timed questions pertaining to information across a document collection. The answer choices for each question included all documents within the collection, and participants selected one or more documents as their final answer.

\textbf{Task 1.} You are a hiring manager for an entry-level technology analyst role at a large financial organization, Acme Inc. You have received resumes from 15 potential candidates. Review the collection of resumes. Your goal is to identify promising candidates to invite for an on-site interview at your offices.
\begin{enumerate}
    \item Candidates must have a strong education background in a relevant technical field. We want to filter all candidates who do not meet this criteria. \textbf{Which candidates DO NOT have a degree in Computer Science, Mathematics, or Engineering?}
    \item Candidates should ideally have experience with at least one programming language and have some prior experience working in the financial sector. \textbf{Which candidates meet BOTH of these criteria?}
    \item The best candidates should be familiar with two processes often used by technology analysts at Acme Inc: 1) statistical data analysis and 2) financial risk analysis. \textbf{Which candidates have demonstrated skills or experience relevant to BOTH of these two processes?}
\end{enumerate}

\textbf{Task 2.} Your financial technology company, Acme Inc., has just relocated to a new office space in San Francisco. You need to hire a cleaning service provider. You have received 10 contract offers from several providers in the area. Review the collection of contracts. Your goal is to identify providers that provide Acme Inc. with good benefits in their contract.
\begin{enumerate}
    \item You are especially interested in ensuring the provider includes carpet cleaning and window cleaning as part of the services offered. \textbf{Which providers DO NOT list BOTH carpet cleaning and window cleaning in their services provided?}
    \item You want to know ahead of time how much you will be paying for each service. You don’t want to pay for services hourly. Instead you want a one-time payment for each service that includes pricing for any required equipment, materials, and tools, as well as fees. Note: Payment-related fees (e.g., late payment, missed payment, overtime) are still acceptable. \textbf{Which providers will bill you a one-time payment per service (including all fees)?}
    \item You haven’t had experience with any of these providers so you want to make sure the contract termination is flexible. In case things don’t work out you want a contract that allows termination at any time with written notice. You might want the ability to take action if the quality of service provided does not meet your needs. \textbf{Which providers have flexible terms for termination AND allow you to take action if service is unsatisfactory?}
\end{enumerate}

\subsection{\system{} Usage Details}
\input{tables/interaction_counts}
Table~\ref{tab:interaction-counts} presents the number of interactions participants had with \system{}'s features throughout the usability study. Participants used different LLM-powered actions to complete their tasks, and often sought to verify the results returned from the LLM by checking the original document context.

\subsection{LLM Prompts} \label{appendix:llm-prompts}

Table \ref{tab:prompts} lists the LLM prompts for the various language understanding services enabling actions within \system{}.

\input{tables/prompts}

%% file: figures/ui_pdf_with_notebook.tex
\begin{figure*}[ht!]
    \centering
    \includegraphics[width=0.9\textwidth]{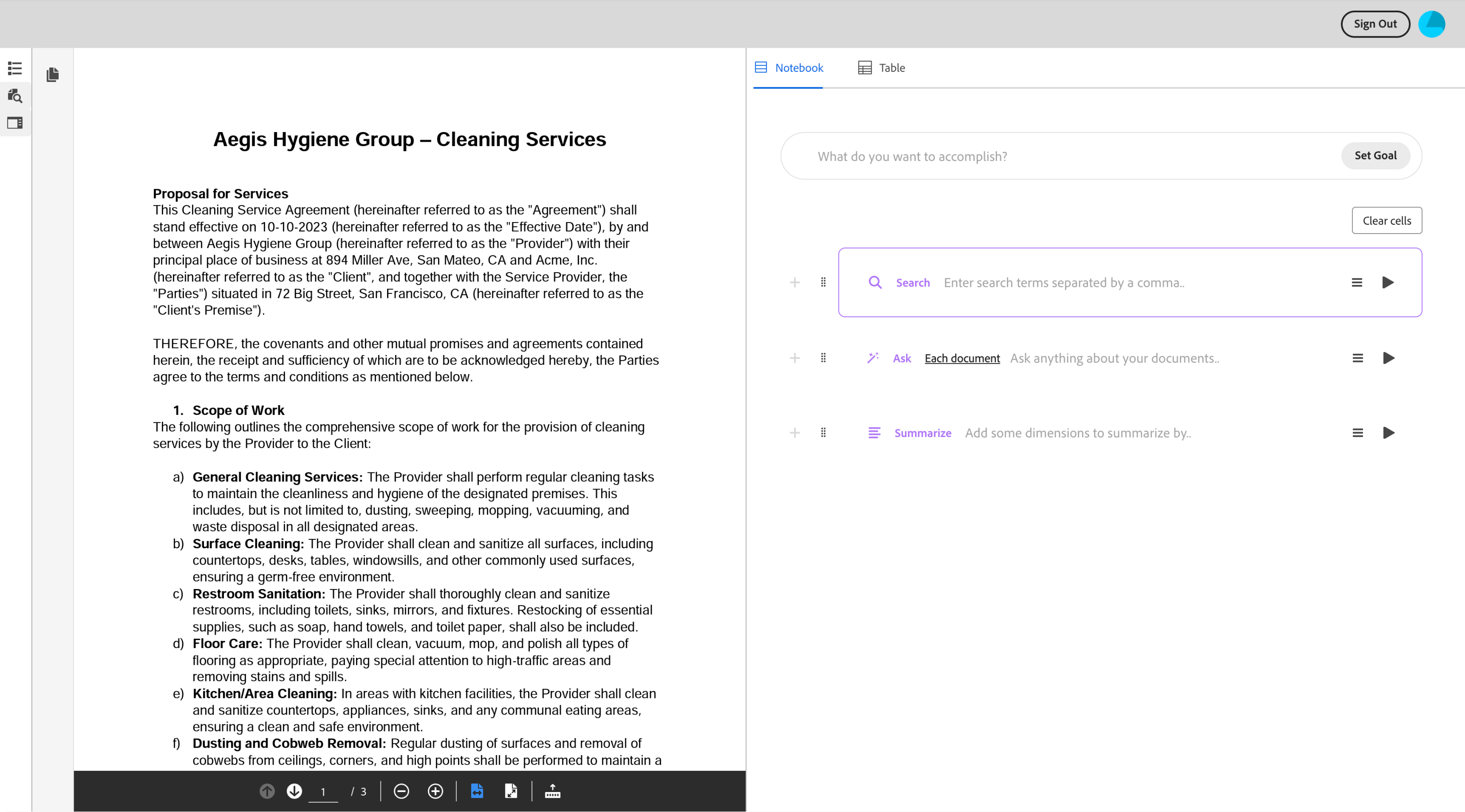}
    \caption{\system{} with both Document and Notebook Views opened side-by-side. This configuration allows users to perform collection-level foraging and sensemaking actions with their notebook, while also verifying details from PDFs on-demand.}
    \Description[Screenshot of Marco’s Document and Notebook Views]{Screenshot of Marco with Document Notebook Views open side-by-side. The Document View shows the start of a cleaning service contract PDF document, and the Notebook VIew shows three created boxes (i.e., cells), Search, Ask Each Document, and Summarize, each outlined in purple and stacked vertically.}
    \label{fig:ui_pdf_with_notebook}
\end{figure*}

%% file: figures/ui_notebook_view.tex
\begin{figure*}[ht!]
    \centering
    \includegraphics[width=0.9\textwidth]{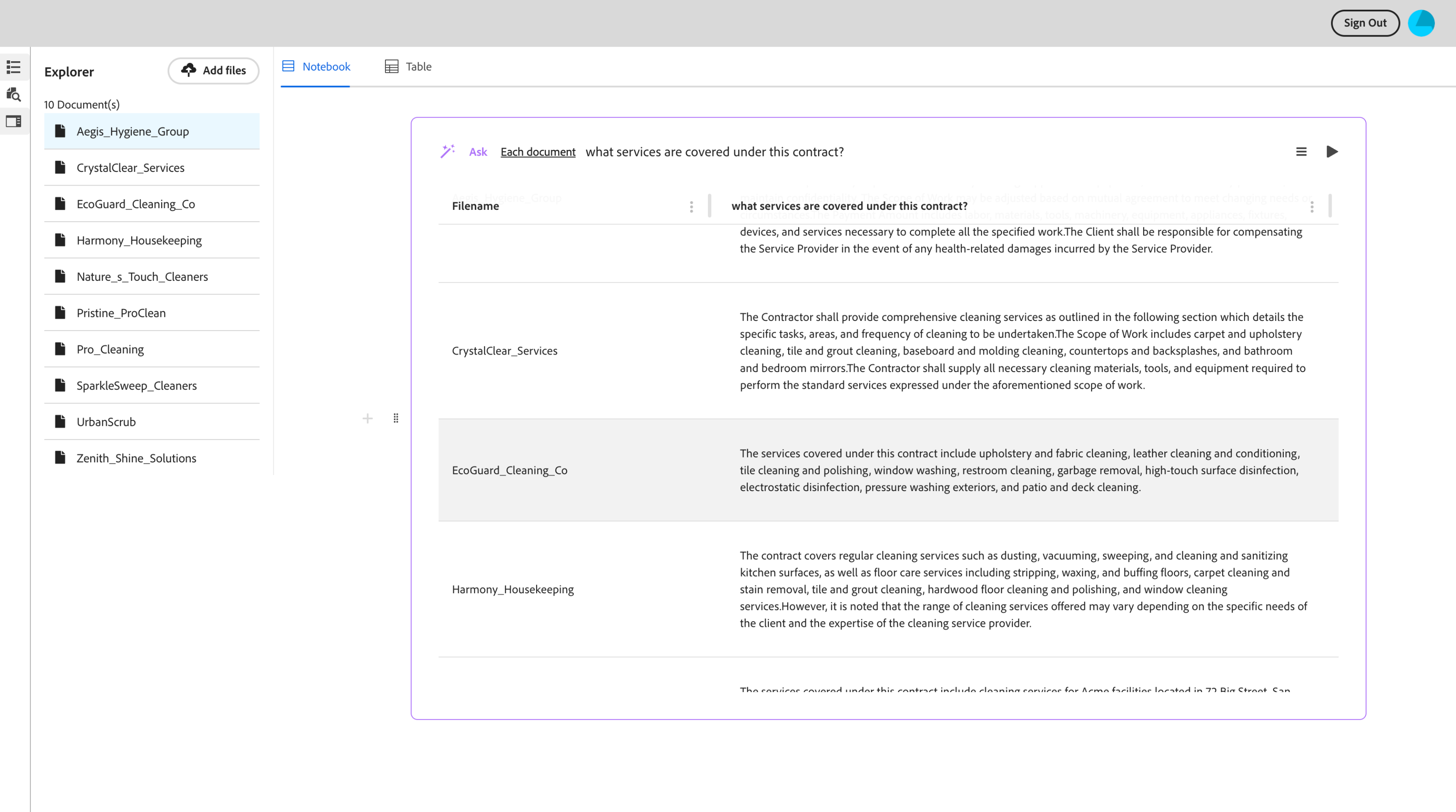}
    \caption{\system{}'s Notebook View. Users primarily interact with \system{} by creating action cells (outlined in purple) to execute information gathering tasks automatically over their entire collection. \system{} encodes execution results as an interactive tabular schema embedded within each individual cell. A left-aligned file manager is also shown, which allows users to see a list of documents in their collection and open individual files, and can be hidden away when unneeded.}
    \Description[Screenshot of Marco’s Notebook Views]{Screenshot of Marco’s Notebook View. A list of files in the collection are shown in a list to the left. The notebook contains one box (cell) outlined in purple. The cell has the text Ask Each Document in purple, and a query text in black: what services are covered under this contract? A result table is shown in the cell with three visible rows and four columns.}
    \label{fig:ui_notebook_view}
\end{figure*}

%% file: figures/ui_table_view.tex
\begin{figure*}[ht!]
    \centering
    \includegraphics[width=0.9\textwidth]{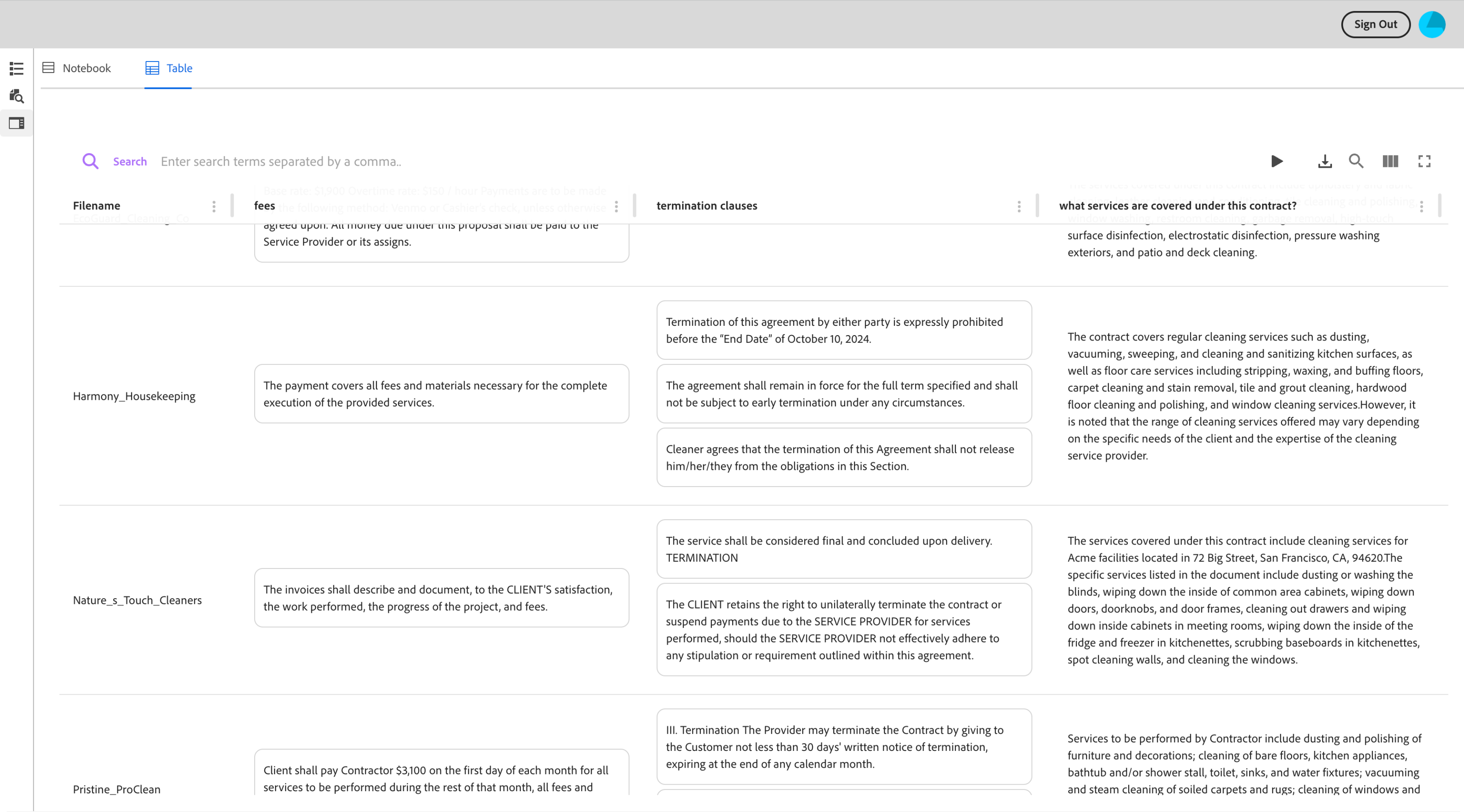}
    \caption{\system{}'s Table View with three executed queries. Users can filter, search, and reorder columns. Users can also add new columns directly from the Table View with an action cell affixed to the table of the table. This action cell executes identically to cells in the Notebook View, but populates the columns in this table directly rather than returning results within an independent cell.}
    \Description[Screenshot of Marco’s Table View]{Screenshot of Marco’s Table View. The view largely consists of a full-screen table with three rows and four columns visible. The first column is the filename, and the remaining three columns are queries issued by a user: fees, termination clauses, and what services are covered under this contract?}
    \label{fig:ui_table_view}
\end{figure*}

%% file: tables/interaction_counts.tex
\renewcommand{\arraystretch}{1.2}
\begin{table*}[ht!]
    \centering
     \Description[Table with Marco interaction counts from the usability study.]{Table with Marco interaction counts from the usability study. Six features are listed on the y-axis, and participant ids are listed on the x-axis. Counts varied from 0 to 28, with most values between 1 and 5.}
    \caption{Counts of participant interactions with \system{} in the usability study. Counts indicate a total across the three task questions completed with \system{} by each participant. \texttt{Search}, \texttt{Ask[Each Document]}, and \texttt{Ask[My Collection]} count the number of times the corresponding action was executed. \texttt{Summarize} was not provided to participants in the study, so no interactions were recorded. \textit{Open PDF - Cell} and \textit{Open PDF - Filename} count the number of times participants opened the PDF in the Document View, either by clicking on a row within a result table in an action cell (context linking) or by clicking on a filename in the file explorer, respectively. \textit{Table View} counts the number of times participants opened the Table View.}
    \begin{tabular}{lrrrrrrrrrrrrrrrr}
    \toprule
     & P1 & P2 & P3 & P4 & P5 & P6 & P7 & P8 & P9 & P10 & P11 & P12 & P13 & P14 & P15 & P16 \\
    \midrule
    \texttt{Search} & 1 & 2 & 4 & 0 & 2 & 3 & 1 & 1 & 3 & 1 & 4 & 3 & 2 & 4 & 4 & 2 \\
    \texttt{Ask[Each Document]} & 2 & 1 & 0 & 0 & 3 & 2 & 4 & 2 & 5 & 1 & 2 & 3 & 3 & 5 & 4 & 1 \\
    \texttt{Ask[My Collection]} & 2 & 0 & 2 & 3 & 2 & 2 & 4 & 1 & 3 & 2 & 2 & 0 & 3 & 0 & 0 & 1 \\
    \textit{Open PDF - Cell} & 1 & 0 & 5 & 0 & 0 & 0 & 1 & 4 & 4 & 8 & 1 & 7 & 4 & 10 & 10 & 2 \\
    \textit{Open PDF - Filename} & 0 & 0 & 1 & 1 & 3 & 0 & 0 & 0 & 0 & 3 & 1 & 3 & 2 & 2 & 4 & 28 \\
    \textit{Table View} & 3 & 0 & 5 & 0 & 2 & 0 & 1 & 1 & 0 & 1 & 1 & 2 & 1 & 3 & 1 & 0 \\
    \bottomrule
    \end{tabular}
    \label{tab:interaction-counts}
\end{table*}

%% file: tables/prompts.tex
\begin{table*}[htbp]
    \centering
    \caption{Prompts used in \system{}. \textbf{\{ \}} refers to a placeholder.}
    \scalebox{0.98}{
    \begin{tabular}{ll}
        \toprule
        \textbf{Task} & \textbf{Prompt} \\
        \midrule
        
        \thead{Search} & \texttt{\thead[l]{
        You will be given excerpts from a document and a search query. Extract the spans \\
        of text that are most relevant to the query, if any, word for word from the \\
        document. Respond with a JSON object with a single key ``snippets'' and a list \\
        of the extracted spans. If there no relevant spans of text, respond with a \\
        JSON object with a single key "snippets" and an empty list. \\\\
        DOCUMENT: \textbf{\{Context\}} \\
        QUERY: Search for ``\textbf{\{Query\}}''
        }}\\
        
        \midrule
        
        \thead{Ask (each document)} & \texttt{\thead[l]{
        You will be given a document and a question. Based on the information contained \\ 
        in the document, answer the question to the best of your abilities. If you \\
        cannot find the answer in the document, respond by saying that the document does \\
        not mention this information. Use the following examples as a guide: \\\\
        \textbf{\{Examples\}} \\\\
        ===================== \\\\
        DOCUMENT: \textbf{\{Context\}} \\
        QUESTION: \textbf{\{Question\}} \\
        ANSWER:
        }}\\
        
        \midrule
        
        \thead{Ask (collection) \\ \textit{Detect missing table info}} & \texttt{\thead[l]{
        Suggest what attributes, if any, are still needed to answer a user's question. \\
        Only suggest a new attribute if a relevant one is not already in the following list. \\
        If no new attribute is needed, just return an empty list. \\\\
        Goal: \textbf{\{Goal\}} \\
        Current attributes: \textbf{\{Columns\}} \\
        Question: \textbf{\{Question\}}
        }} \\

        \thead{Ask (collection) \\ \textit{Synthesize response}} & \texttt{\thead[l]{
        Answer the user's question based on information in the following table. The \\
        table contains information extracted directly from a collection of documents \\
        which may be relevant to the user's query. When possible, try to provide a \\
        concise explanation for your answer based on the information provided in the \\
        table. If the question cannot be answered given the provided information, \\
        respond ``I don't know''. \\\\
        Table: \textbf{\{Table\}} \\
        Question: \textbf{\{Question\}} \\
        Answer:
        }} \\

        \midrule

        \thead{AI Suggestions} & \texttt{\thead[l]{
        You are a intelligent document reading assistant helping users explore and \\
        understand their documents to achieve their goals (\textbf{\{Goal\}}). \\\\
        The following shows examples of the types of documents the user is working with \\
        (but are not the exact documents they have). \\\\
        Sample document: \textbf{\{Up to 1000 characters of Sample Document 1\}} \\
        Sample document: \textbf{\{Up to 1000 characters of Sample Document 2\}} \\
        Sample document: \textbf{\{Up to 1000 characters of Sample Document 3\}} \\\\
        The user has previously searched for the following things within their documents: \textbf{\{Searches\}} \\\\
        The user has previously asked the following questions of their documents: \textbf{\{Questions\}} \\\\
        Suggest up to two other searches and questions they could ask their documents. \\
        These search queries and questions should be answerable given the document and nothing else. \\
        Respond with exactly one JSON object with two keys: ``suggested\_searches'' and ``suggested\_questions''. \\
        If no relevant searches or questions can be asked, respond with an empty list for both \\
        ``suggested\_searches'' and ``suggested\_questions''.
        }} \\
        
        \bottomrule
    \end{tabular}
    }
    \label{tab:prompts}
\end{table*}

%% file: main.bbl

\begin{thebibliography}{82}


\ifx \showCODEN    \undefined \def \showCODEN     #1{\unskip}     \fi
\ifx \showDOI      \undefined \def \showDOI       #1{#1}\fi
\ifx \showISBNx    \undefined \def \showISBNx     #1{\unskip}     \fi
\ifx \showISBNxiii \undefined \def \showISBNxiii  #1{\unskip}     \fi
\ifx \showISSN     \undefined \def \showISSN      #1{\unskip}     \fi
\ifx \showLCCN     \undefined \def \showLCCN      #1{\unskip}     \fi
\ifx \shownote     \undefined \def \shownote      #1{#1}          \fi
\ifx \showarticletitle \undefined \def \showarticletitle #1{#1}   \fi
\ifx \showURL      \undefined \def \showURL       {\relax}        \fi
\providecommand\bibfield[2]{#2}
\providecommand\bibinfo[2]{#2}
\providecommand\natexlab[1]{#1}
\providecommand\showeprint[2][]{arXiv:#2}

\bibitem[Adler et~al\mbox{.}(1998)]%
        {adler1998diary}
\bibfield{author}{\bibinfo{person}{Annette Adler}, \bibinfo{person}{Anuj
  Gujar}, \bibinfo{person}{Beverly~L. Harrison}, \bibinfo{person}{Kenton
  O'Hara}, {and} \bibinfo{person}{Abigail Sellen}.}
  \bibinfo{year}{1998}\natexlab{}.
\newblock \showarticletitle{A Diary Study of Work-Related Reading: Design
  Implications for Digital Reading Devices}. In
  \bibinfo{booktitle}{\emph{Proceedings of the SIGCHI Conference on Human
  Factors in Computing Systems}} (Los Angeles, California, USA)
  \emph{(\bibinfo{series}{CHI '98})}. \bibinfo{publisher}{ACM
  Press/Addison-Wesley Publishing Co.}, \bibinfo{address}{USA},
  \bibinfo{pages}{241–248}.
\newblock
\urldef\tempurl%
\url{https://doi.org/10.1145/274644.274679}
\showURL{%
\tempurl}


\bibitem[Adobe(2023)]%
        {pdfExtractAPI}
\bibfield{author}{\bibinfo{person}{Adobe}.} \bibinfo{year}{2023}\natexlab{}.
\newblock \bibinfo{title}{PDF Extract API}.
\newblock
\newblock
\urldef\tempurl%
\url{https://developer.adobe.com/document-services/docs/overview/pdf-extract-api/}
\showURL{%
Retrieved August 27, 2023 from \tempurl}


\bibitem[Albert and De~Ruiter(2018)]%
        {albert2018repair}
\bibfield{author}{\bibinfo{person}{Saul Albert} {and} \bibinfo{person}{Jan~P
  De~Ruiter}.} \bibinfo{year}{2018}\natexlab{}.
\newblock \showarticletitle{Repair: the interface between interaction and
  cognition}.
\newblock \bibinfo{journal}{\emph{Topics in cognitive science}}
  \bibinfo{volume}{10}, \bibinfo{number}{2} (\bibinfo{year}{2018}),
  \bibinfo{pages}{279--313}.
\newblock


\bibitem[Amershi et~al\mbox{.}(2019)]%
        {amershi_guidelines_2019}
\bibfield{author}{\bibinfo{person}{Saleema Amershi}, \bibinfo{person}{Dan
  Weld}, \bibinfo{person}{Mihaela Vorvoreanu}, \bibinfo{person}{Adam Fourney},
  \bibinfo{person}{Besmira Nushi}, \bibinfo{person}{Penny Collisson},
  \bibinfo{person}{Jina Suh}, \bibinfo{person}{Shamsi Iqbal},
  \bibinfo{person}{Paul~N. Bennett}, \bibinfo{person}{Kori Inkpen},
  \bibinfo{person}{Jaime Teevan}, \bibinfo{person}{Ruth Kikin-Gil}, {and}
  \bibinfo{person}{Eric Horvitz}.} \bibinfo{year}{2019}\natexlab{}.
\newblock \showarticletitle{Guidelines for Human-AI Interaction}. In
  \bibinfo{booktitle}{\emph{Proceedings of the 2019 CHI Conference on Human
  Factors in Computing Systems}}. \bibinfo{publisher}{ACM},
  \bibinfo{address}{Glasgow Scotland Uk}, \bibinfo{pages}{1--13}.
\newblock
\urldef\tempurl%
\url{https://dl.acm.org/doi/10.1145/3290605.3300233}
\showURL{%
\tempurl}


\bibitem[Ashktorab et~al\mbox{.}(2021)]%
        {ashktorab2021ai}
\bibfield{author}{\bibinfo{person}{Zahra Ashktorab}, \bibinfo{person}{Michael
  Desmond}, \bibinfo{person}{Josh Andres}, \bibinfo{person}{Michael Muller},
  \bibinfo{person}{Narendra~Nath Joshi}, \bibinfo{person}{Michelle Brachman},
  \bibinfo{person}{Aabhas Sharma}, \bibinfo{person}{Kristina Brimijoin},
  \bibinfo{person}{Qian Pan}, \bibinfo{person}{Christine~T Wolf},
  {et~al\mbox{.}}} \bibinfo{year}{2021}\natexlab{}.
\newblock \showarticletitle{Ai-assisted human labeling: Batching for efficiency
  without overreliance}.
\newblock \bibinfo{journal}{\emph{Proceedings of the ACM on Human-Computer
  Interaction}} \bibinfo{volume}{5}, \bibinfo{number}{CSCW1}
  (\bibinfo{year}{2021}), \bibinfo{pages}{1--27}.
\newblock


\bibitem[Baldonado and Winograd(1997)]%
        {baldonado_sensemaker_1997}
\bibfield{author}{\bibinfo{person}{Michelle Q.~Wang Baldonado} {and}
  \bibinfo{person}{Terry Winograd}.} \bibinfo{year}{1997}\natexlab{}.
\newblock \showarticletitle{SenseMaker: an information-exploration interface
  supporting the contextual evolution of a user's interests}. In
  \bibinfo{booktitle}{\emph{Proceedings of the ACM SIGCHI Conference on Human
  factors in computing systems}}. \bibinfo{publisher}{ACM},
  \bibinfo{address}{Atlanta Georgia USA}, \bibinfo{pages}{11--18}.
\newblock
\urldef\tempurl%
\url{https://dl.acm.org/doi/10.1145/258549.258563}
\showURL{%
\tempurl}


\bibitem[Bang et~al\mbox{.}(2023)]%
        {bang_multitask_2023}
\bibfield{author}{\bibinfo{person}{Yejin Bang}, \bibinfo{person}{Samuel
  Cahyawijaya}, \bibinfo{person}{Nayeon Lee}, \bibinfo{person}{Wenliang Dai},
  \bibinfo{person}{Dan Su}, \bibinfo{person}{Bryan Wilie},
  \bibinfo{person}{Holy Lovenia}, \bibinfo{person}{Ziwei Ji},
  \bibinfo{person}{Tiezheng Yu}, \bibinfo{person}{Willy Chung},
  \bibinfo{person}{Quyet~V. Do}, \bibinfo{person}{Yan Xu}, {and}
  \bibinfo{person}{Pascale Fung}.} \bibinfo{year}{2023}\natexlab{}.
\newblock \bibinfo{title}{A Multitask, Multilingual, Multimodal Evaluation of
  ChatGPT on Reasoning, Hallucination, and Interactivity}.
\newblock
\newblock
\urldef\tempurl%
\url{http://arxiv.org/abs/2302.04023}
\showURL{%
\tempurl}


\bibitem[Bates et~al\mbox{.}(2015)]%
        {bates_fitting_2015}
\bibfield{author}{\bibinfo{person}{Douglas Bates}, \bibinfo{person}{Martin
  Mächler}, \bibinfo{person}{Ben Bolker}, {and} \bibinfo{person}{Steve
  Walker}.} \bibinfo{year}{2015}\natexlab{}.
\newblock \showarticletitle{Fitting Linear Mixed-Effects Models Using lme4}.
\newblock \bibinfo{journal}{\emph{Journal of Statistical Software}}
  \bibinfo{volume}{67}, \bibinfo{number}{1} (\bibinfo{year}{2015}),
  \bibinfo{pages}{1--48}.
\newblock
\urldef\tempurl%
\url{https://www.jstatsoft.org/index.php/jss/article/view/v067i01}
\showURL{%
\tempurl}


\bibitem[Boyatzis(1998)]%
        {boyatzis1998transforming}
\bibfield{author}{\bibinfo{person}{Richard~E Boyatzis}.}
  \bibinfo{year}{1998}\natexlab{}.
\newblock \bibinfo{booktitle}{\emph{Transforming qualitative information:
  Thematic analysis and code development}}.
\newblock \bibinfo{publisher}{SAGE}, \bibinfo{address}{Thousand Oaks, CA}.
\newblock


\bibitem[Braun and Clarke(2006)]%
        {braun2006using}
\bibfield{author}{\bibinfo{person}{Virginia Braun} {and}
  \bibinfo{person}{Victoria Clarke}.} \bibinfo{year}{2006}\natexlab{}.
\newblock \showarticletitle{Using thematic analysis in psychology}.
\newblock \bibinfo{journal}{\emph{Qualitative research in psychology}}
  \bibinfo{volume}{3}, \bibinfo{number}{2} (\bibinfo{year}{2006}),
  \bibinfo{pages}{77--101}.
\newblock


\bibitem[Brooke(1996)]%
        {brooke1996sus}
\bibfield{author}{\bibinfo{person}{John Brooke}.}
  \bibinfo{year}{1996}\natexlab{}.
\newblock \showarticletitle{SUS: A Quick and Dirty Usability Scale}.
\newblock \bibinfo{journal}{\emph{Usability Evaluation in Industry}}
  \bibinfo{volume}{189}, \bibinfo{number}{3} (\bibinfo{year}{1996}),
  \bibinfo{pages}{189--194}.
\newblock


\bibitem[Cambon et~al\mbox{.}(2023)]%
        {cambon2023early}
\bibfield{author}{\bibinfo{person}{Alexia Cambon}, \bibinfo{person}{Brent
  Hecht}, \bibinfo{person}{Ben Edelman}, \bibinfo{person}{Donald Ngwe},
  \bibinfo{person}{Sonia Jaffe}, \bibinfo{person}{Amy Heger},
  \bibinfo{person}{Mihaela Vorvoreanu}, \bibinfo{person}{Sida Peng},
  \bibinfo{person}{Jake Hofman}, \bibinfo{person}{Alex Farach},
  \bibinfo{person}{Margarita Bermejo-Cano}, \bibinfo{person}{Eric Knudsen},
  \bibinfo{person}{James Bono}, \bibinfo{person}{Hardik Sanghavi},
  \bibinfo{person}{Sofia Spatharioti}, \bibinfo{person}{David Rothschild},
  \bibinfo{person}{Daniel~G. Goldstein}, \bibinfo{person}{Eirini Kalliamvakou},
  \bibinfo{person}{Peter Cihon}, \bibinfo{person}{Mert Demirer},
  \bibinfo{person}{Michael Schwarz}, {and} \bibinfo{person}{Jaime Teevan}.}
  \bibinfo{year}{2023}\natexlab{}.
\newblock \bibinfo{booktitle}{\emph{Early LLM-based Tools for Enterprise
  Information Workers Likely Provide Meaningful Boosts to Productivity}}.
\newblock \bibinfo{type}{{T}echnical {R}eport} MSR-TR-2023-43.
  \bibinfo{institution}{Microsoft}.
\newblock
\urldef\tempurl%
\url{https://www.microsoft.com/en-us/research/publication/early-llm-based-tools-for-enterprise-information-workers-likely-provide-meaningful-boosts-to-productivity/}
\showURL{%
\tempurl}


\bibitem[Chang et~al\mbox{.}(2020)]%
        {chang_mesh_2020}
\bibfield{author}{\bibinfo{person}{Joseph~Chee Chang}, \bibinfo{person}{Nathan
  Hahn}, {and} \bibinfo{person}{Aniket Kittur}.}
  \bibinfo{year}{2020}\natexlab{}.
\newblock \showarticletitle{Mesh: Scaffolding Comparison Tables for Online
  Decision Making}. In \bibinfo{booktitle}{\emph{Proceedings of the 33rd Annual
  ACM Symposium on User Interface Software and Technology}}.
  \bibinfo{publisher}{ACM}, \bibinfo{address}{Virtual Event USA},
  \bibinfo{pages}{391--405}.
\newblock
\urldef\tempurl%
\url{https://dl.acm.org/doi/10.1145/3379337.3415865}
\showURL{%
\tempurl}


\bibitem[Chang et~al\mbox{.}(2019)]%
        {chang_searchlens_2019}
\bibfield{author}{\bibinfo{person}{Joseph~Chee Chang}, \bibinfo{person}{Nathan
  Hahn}, \bibinfo{person}{Adam Perer}, {and} \bibinfo{person}{Aniket Kittur}.}
  \bibinfo{year}{2019}\natexlab{}.
\newblock \showarticletitle{SearchLens: composing and capturing complex user
  interests for exploratory search}. In \bibinfo{booktitle}{\emph{Proceedings
  of the 24th International Conference on Intelligent User Interfaces}}.
  \bibinfo{publisher}{ACM}, \bibinfo{address}{Marina del Ray California},
  \bibinfo{pages}{498--509}.
\newblock
\urldef\tempurl%
\url{https://dl.acm.org/doi/10.1145/3301275.3302321}
\showURL{%
\tempurl}


\bibitem[Chen and Wang(2017)]%
        {chen_explaining_2017}
\bibfield{author}{\bibinfo{person}{Li Chen} {and} \bibinfo{person}{Feng Wang}.}
  \bibinfo{year}{2017}\natexlab{}.
\newblock \showarticletitle{Explaining Recommendations Based on Feature
  Sentiments in Product Reviews}. In \bibinfo{booktitle}{\emph{Proceedings of
  the 22nd International Conference on Intelligent User Interfaces}}
  \emph{(\bibinfo{series}{IUI '17})}. \bibinfo{publisher}{Association for
  Computing Machinery}, \bibinfo{address}{New York, NY, USA},
  \bibinfo{pages}{17--28}.
\newblock
\urldef\tempurl%
\url{https://dl.acm.org/doi/10.1145/3025171.3025173}
\showURL{%
\tempurl}


\bibitem[Chen et~al\mbox{.}(2014)]%
        {chen_experiment_2014}
\bibfield{author}{\bibinfo{person}{Li Chen}, \bibinfo{person}{Feng Wang},
  \bibinfo{person}{Luole Qi}, {and} \bibinfo{person}{Fengfeng Liang}.}
  \bibinfo{year}{2014}\natexlab{}.
\newblock \showarticletitle{Experiment on sentiment embedded comparison
  interface}.
\newblock \bibinfo{journal}{\emph{Knowledge-Based Systems}}
  \bibinfo{volume}{64} (\bibinfo{date}{July} \bibinfo{year}{2014}),
  \bibinfo{pages}{44--58}.
\newblock
\urldef\tempurl%
\url{https://www.sciencedirect.com/science/article/pii/S0950705114001099}
\showURL{%
\tempurl}


\bibitem[Chen and Xia(2022)]%
        {chen2022crossdata}
\bibfield{author}{\bibinfo{person}{Zhutian Chen} {and} \bibinfo{person}{Haijun
  Xia}.} \bibinfo{year}{2022}\natexlab{}.
\newblock \showarticletitle{CrossData: Leveraging Text-Data Connections for
  Authoring Data Documents}. In \bibinfo{booktitle}{\emph{Proceedings of the
  2022 CHI Conference on Human Factors in Computing Systems}} (New Orleans, LA,
  USA) \emph{(\bibinfo{series}{CHI '22})}. \bibinfo{publisher}{Association for
  Computing Machinery}, \bibinfo{address}{New York, NY, USA}, Article
  \bibinfo{articleno}{95}, \bibinfo{numpages}{15}~pages.
\newblock
\showISBNx{9781450391573}
\urldef\tempurl%
\url{https://doi.org/10.1145/3491102.3517485}
\showDOI{\tempurl}


\bibitem[Cheng et~al\mbox{.}(2019)]%
        {cheng2019explaining}
\bibfield{author}{\bibinfo{person}{Hao-Fei Cheng}, \bibinfo{person}{Ruotong
  Wang}, \bibinfo{person}{Zheng Zhang}, \bibinfo{person}{Fiona O'Connell},
  \bibinfo{person}{Terrance Gray}, \bibinfo{person}{F.~Maxwell Harper}, {and}
  \bibinfo{person}{Haiyi Zhu}.} \bibinfo{year}{2019}\natexlab{}.
\newblock \showarticletitle{Explaining Decision-Making Algorithms through UI:
  Strategies to Help Non-Expert Stakeholders}. In
  \bibinfo{booktitle}{\emph{Proceedings of the 2019 CHI Conference on Human
  Factors in Computing Systems}} (Glasgow, Scotland Uk)
  \emph{(\bibinfo{series}{CHI '19})}. \bibinfo{publisher}{Association for
  Computing Machinery}, \bibinfo{address}{New York, NY, USA},
  \bibinfo{pages}{1–12}.
\newblock
\showISBNx{9781450359702}
\urldef\tempurl%
\url{https://doi.org/10.1145/3290605.3300789}
\showDOI{\tempurl}


\bibitem[Cnaan et~al\mbox{.}(1997)]%
        {cnaan1997using}
\bibfield{author}{\bibinfo{person}{Avital Cnaan}, \bibinfo{person}{Nan~M
  Laird}, {and} \bibinfo{person}{Peter Slasor}.}
  \bibinfo{year}{1997}\natexlab{}.
\newblock \showarticletitle{Using the general linear mixed model to analyse
  unbalanced repeated measures and longitudinal data}.
\newblock \bibinfo{journal}{\emph{Statistics in medicine}}
  \bibinfo{volume}{16}, \bibinfo{number}{20} (\bibinfo{year}{1997}),
  \bibinfo{pages}{2349--2380}.
\newblock


\bibitem[Cudeck(1996)]%
        {cudeck1996mixed}
\bibfield{author}{\bibinfo{person}{Robert Cudeck}.}
  \bibinfo{year}{1996}\natexlab{}.
\newblock \showarticletitle{Mixed-effects models in the study of individual
  differences with repeated measures data}.
\newblock \bibinfo{journal}{\emph{Multivariate behavioral research}}
  \bibinfo{volume}{31}, \bibinfo{number}{3} (\bibinfo{year}{1996}),
  \bibinfo{pages}{371--403}.
\newblock


\bibitem[Cutting et~al\mbox{.}(1992)]%
        {cutting_scattergather_1992}
\bibfield{author}{\bibinfo{person}{Douglass~R. Cutting},
  \bibinfo{person}{David~R. Karger}, \bibinfo{person}{Jan~O. Pedersen}, {and}
  \bibinfo{person}{John~W. Tukey}.} \bibinfo{year}{1992}\natexlab{}.
\newblock \showarticletitle{Scatter/Gather: a cluster-based approach to
  browsing large document collections}. In
  \bibinfo{booktitle}{\emph{Proceedings of the 15th annual international ACM
  SIGIR conference on Research and development in information retrieval}}
  \emph{(\bibinfo{series}{SIGIR '92})}. \bibinfo{publisher}{Association for
  Computing Machinery}, \bibinfo{address}{New York, NY, USA},
  \bibinfo{pages}{318--329}.
\newblock
\urldef\tempurl%
\url{https://dl.acm.org/doi/10.1145/133160.133214}
\showURL{%
\tempurl}


\bibitem[Dell'Acqua et~al\mbox{.}(2023)]%
        {dell2023navigating}
\bibfield{author}{\bibinfo{person}{Fabrizio Dell'Acqua},
  \bibinfo{person}{Edward McFowland}, \bibinfo{person}{Ethan~R Mollick},
  \bibinfo{person}{Hila Lifshitz-Assaf}, \bibinfo{person}{Katherine Kellogg},
  \bibinfo{person}{Saran Rajendran}, \bibinfo{person}{Lisa Krayer},
  \bibinfo{person}{Fran{\c{c}}ois Candelon}, {and} \bibinfo{person}{Karim~R
  Lakhani}.} \bibinfo{year}{2023}\natexlab{}.
\newblock \showarticletitle{Navigating the jagged technological frontier: field
  experimental evidence of the effects of AI on knowledge worker productivity
  and quality}.
\newblock \bibinfo{journal}{\emph{Harvard Business School Technology \&
  Operations Mgt. Unit Working Paper}} \bibinfo{number}{24-013}
  (\bibinfo{year}{2023}).
\newblock


\bibitem[Do et~al\mbox{.}(2023)]%
        {do_err_2023}
\bibfield{author}{\bibinfo{person}{Hyo~Jin Do}, \bibinfo{person}{Ha-Kyung
  Kong}, \bibinfo{person}{Pooja Tetali}, \bibinfo{person}{Jaewook Lee}, {and}
  \bibinfo{person}{Brian~P. Bailey}.} \bibinfo{year}{2023}\natexlab{}.
\newblock \showarticletitle{To Err is AI: Imperfect Interventions and Repair in
  a Conversational Agent Facilitating Group Chat Discussions}.
\newblock \bibinfo{journal}{\emph{Proceedings of the ACM on Human-Computer
  Interaction}} \bibinfo{volume}{7}, \bibinfo{number}{CSCW1}
  (\bibinfo{date}{April} \bibinfo{year}{2023}), \bibinfo{pages}{1--23}.
\newblock
\urldef\tempurl%
\url{https://dl.acm.org/doi/10.1145/3579532}
\showURL{%
\tempurl}


\bibitem[Dontcheva et~al\mbox{.}(2007)]%
        {dontcheva_relations_2007}
\bibfield{author}{\bibinfo{person}{Mira Dontcheva}, \bibinfo{person}{Steven~M.
  Drucker}, \bibinfo{person}{David Salesin}, {and} \bibinfo{person}{Michael~F.
  Cohen}.} \bibinfo{year}{2007}\natexlab{}.
\newblock \showarticletitle{Relations, Cards, and Search Templates: User-Guided
  Web Data Integration and Layout}. In \bibinfo{booktitle}{\emph{Proceedings of
  the 20th Annual ACM Symposium on User Interface Software and Technology}}
  (Newport, Rhode Island, USA) \emph{(\bibinfo{series}{UIST '07})}.
  \bibinfo{publisher}{Association for Computing Machinery},
  \bibinfo{address}{New York, NY, USA}, \bibinfo{pages}{61–70}.
\newblock
\urldef\tempurl%
\url{https://doi.org/10.1145/1294211.1294224}
\showURL{%
\tempurl}


\bibitem[English et~al\mbox{.}(2002)]%
        {english_hierarchical_2002}
\bibfield{author}{\bibinfo{person}{Jennifer English}, \bibinfo{person}{Marti
  Hearst}, \bibinfo{person}{Rashmi Sinha}, \bibinfo{person}{Kirsten
  Swearingen}, {and} \bibinfo{person}{Ka-Ping Yee}.}
  \bibinfo{year}{2002}\natexlab{}.
\newblock \showarticletitle{Hierarchical Faceted Metadata in Site Search
  Interfaces}. In \bibinfo{booktitle}{\emph{CHI '02 Extended Abstracts on Human
  Factors in Computing Systems}} (Minneapolis, Minnesota, USA)
  \emph{(\bibinfo{series}{CHI EA '02})}. \bibinfo{publisher}{Association for
  Computing Machinery}, \bibinfo{address}{New York, NY, USA},
  \bibinfo{pages}{628–639}.
\newblock
\urldef\tempurl%
\url{https://doi.org/10.1145/506443.506517}
\showURL{%
\tempurl}


\bibitem[Felix et~al\mbox{.}(2018)]%
        {felix2018exploratory}
\bibfield{author}{\bibinfo{person}{Cristian Felix}, \bibinfo{person}{Aritra
  Dasgupta}, {and} \bibinfo{person}{Enrico Bertini}.}
  \bibinfo{year}{2018}\natexlab{}.
\newblock \showarticletitle{The Exploratory Labeling Assistant:
  Mixed-Initiative Label Curation with Large Document Collections}. In
  \bibinfo{booktitle}{\emph{Proceedings of the 31st Annual ACM Symposium on
  User Interface Software and Technology}} (Berlin, Germany)
  \emph{(\bibinfo{series}{UIST '18})}. \bibinfo{publisher}{Association for
  Computing Machinery}, \bibinfo{address}{New York, NY, USA},
  \bibinfo{pages}{153–164}.
\newblock
\showISBNx{9781450359481}
\urldef\tempurl%
\url{https://doi.org/10.1145/3242587.3242596}
\showDOI{\tempurl}


\bibitem[Ferreira et~al\mbox{.}(2019)]%
        {ferreira2019should}
\bibfield{author}{\bibinfo{person}{Juliana~Jansen Ferreira},
  \bibinfo{person}{Ana Fucs}, {and} \bibinfo{person}{Vin\'{\i}cius Segura}.}
  \bibinfo{year}{2019}\natexlab{}.
\newblock \showarticletitle{Should I Interfere' AI-Assistants' Interaction with
  Knowledge Workers: A Case Study in the Oil and Gas Industry}. In
  \bibinfo{booktitle}{\emph{Extended Abstracts of the 2019 CHI Conference on
  Human Factors in Computing Systems}} (Glasgow, Scotland Uk)
  \emph{(\bibinfo{series}{CHI EA '19})}. \bibinfo{publisher}{Association for
  Computing Machinery}, \bibinfo{address}{New York, NY, USA},
  \bibinfo{pages}{1–7}.
\newblock
\urldef\tempurl%
\url{https://doi.org/10.1145/3290607.3299052}
\showURL{%
\tempurl}


\bibitem[Fok and Weld(2023)]%
        {fok_search_2023}
\bibfield{author}{\bibinfo{person}{Raymond Fok} {and}
  \bibinfo{person}{Daniel~S. Weld}.} \bibinfo{year}{2023}\natexlab{}.
\newblock \bibinfo{title}{In Search of Verifiability: Explanations Rarely
  Enable Complementary Performance in AI-Advised Decision Making}.
\newblock
\newblock


\bibitem[Google(2023)]%
        {bard}
\bibfield{author}{\bibinfo{person}{Google}.} \bibinfo{year}{2023}\natexlab{}.
\newblock \bibinfo{title}{Bard}.
\newblock
\newblock
\urldef\tempurl%
\url{https://bard.google.com/}
\showURL{%
Retrieved December 4, 2023 from \tempurl}


\bibitem[Hahn et~al\mbox{.}(2018)]%
        {hahn_bento_2018}
\bibfield{author}{\bibinfo{person}{Nathan Hahn}, \bibinfo{person}{Joseph~Chee
  Chang}, {and} \bibinfo{person}{Aniket Kittur}.}
  \bibinfo{year}{2018}\natexlab{}.
\newblock \showarticletitle{Bento Browser: Complex Mobile Search Without Tabs}.
  In \bibinfo{booktitle}{\emph{Proceedings of the 2018 CHI Conference on Human
  Factors in Computing Systems}} \emph{(\bibinfo{series}{CHI '18})}.
  \bibinfo{publisher}{Association for Computing Machinery},
  \bibinfo{address}{New York, NY, USA}, \bibinfo{pages}{1--12}.
\newblock
\urldef\tempurl%
\url{https://dl.acm.org/doi/10.1145/3173574.3173825}
\showURL{%
\tempurl}


\bibitem[Han et~al\mbox{.}(2020)]%
        {han2020textlets}
\bibfield{author}{\bibinfo{person}{Han~L. Han}, \bibinfo{person}{Miguel~A.
  Renom}, \bibinfo{person}{Wendy~E. Mackay}, {and} \bibinfo{person}{Michel
  Beaudouin-Lafon}.} \bibinfo{year}{2020}\natexlab{}.
\newblock \showarticletitle{Textlets: Supporting Constraints and Consistency in
  Text Documents}. In \bibinfo{booktitle}{\emph{Proceedings of the 2020 CHI
  Conference on Human Factors in Computing Systems}} (Honolulu, HI, USA)
  \emph{(\bibinfo{series}{CHI '20})}. \bibinfo{publisher}{Association for
  Computing Machinery}, \bibinfo{address}{New York, NY, USA},
  \bibinfo{pages}{1–13}.
\newblock
\urldef\tempurl%
\url{https://doi.org/10.1145/3313831.3376804}
\showURL{%
\tempurl}


\bibitem[Han et~al\mbox{.}(2022)]%
        {han_passages_2022}
\bibfield{author}{\bibinfo{person}{Han~L. Han}, \bibinfo{person}{Junhang Yu},
  \bibinfo{person}{Raphael Bournet}, \bibinfo{person}{Alexandre Ciorascu},
  \bibinfo{person}{Wendy~E. Mackay}, {and} \bibinfo{person}{Michel
  Beaudouin-Lafon}.} \bibinfo{year}{2022}\natexlab{}.
\newblock \showarticletitle{Passages: Interacting with Text Across Documents}.
  In \bibinfo{booktitle}{\emph{CHI Conference on Human Factors in Computing
  Systems}}. \bibinfo{publisher}{ACM}, \bibinfo{address}{New Orleans LA USA},
  \bibinfo{pages}{1--17}.
\newblock
\urldef\tempurl%
\url{https://dl.acm.org/doi/10.1145/3491102.3502052}
\showURL{%
\tempurl}


\bibitem[Hearst and Degler(2013)]%
        {hearst_sewing_2013}
\bibfield{author}{\bibinfo{person}{Marti~A. Hearst} {and}
  \bibinfo{person}{Duane Degler}.} \bibinfo{year}{2013}\natexlab{}.
\newblock \showarticletitle{Sewing the Seams of Sensemaking: A Practical
  Interface for Tagging and Organizing Saved Search Results}. In
  \bibinfo{booktitle}{\emph{Proceedings of the Symposium on Human-Computer
  Interaction and Information Retrieval}} \emph{(\bibinfo{series}{HCIR '13})}.
  \bibinfo{publisher}{Association for Computing Machinery},
  \bibinfo{address}{New York, NY, USA}, \bibinfo{pages}{1--10}.
\newblock
\urldef\tempurl%
\url{https://dl.acm.org/doi/10.1145/2528394.2528398}
\showURL{%
\tempurl}


\bibitem[Hearst and Stoica(2009)]%
        {hearst_nlp_2009}
\bibfield{author}{\bibinfo{person}{Marti~A. Hearst} {and}
  \bibinfo{person}{Emilia Stoica}.} \bibinfo{year}{2009}\natexlab{}.
\newblock \showarticletitle{NLP support for faceted navigation in scholarly
  collections}. In \bibinfo{booktitle}{\emph{Proceedings of the 2009 Workshop
  on Text and Citation Analysis for Scholarly Digital Libraries - NLPIR4DL
  '09}}. \bibinfo{publisher}{Association for Computational Linguistics},
  \bibinfo{address}{Suntec, Singapore}, \bibinfo{pages}{62}.
\newblock
\urldef\tempurl%
\url{http://portal.acm.org/citation.cfm?doid=1699750.1699760}
\showURL{%
\tempurl}


\bibitem[Heidary~Dahooie et~al\mbox{.}(2011)]%
        {heidary2011activity}
\bibfield{author}{\bibinfo{person}{Jalil Heidary~Dahooie},
  \bibinfo{person}{Abbas Afrazeh}, {and} \bibinfo{person}{Seyed Mohammad
  Moathar~Hosseini}.} \bibinfo{year}{2011}\natexlab{}.
\newblock \showarticletitle{An activity-based framework for quantification of
  knowledge work}.
\newblock \bibinfo{journal}{\emph{Journal of Knowledge Management}}
  \bibinfo{volume}{15}, \bibinfo{number}{3} (\bibinfo{year}{2011}),
  \bibinfo{pages}{422--444}.
\newblock


\bibitem[Horvitz(1999)]%
        {horvitz_principles_1999}
\bibfield{author}{\bibinfo{person}{Eric Horvitz}.}
  \bibinfo{year}{1999}\natexlab{}.
\newblock \showarticletitle{Principles of mixed-initiative user interfaces}. In
  \bibinfo{booktitle}{\emph{Proceedings of the SIGCHI conference on Human
  Factors in Computing Systems}} \emph{(\bibinfo{series}{CHI '99})}.
  \bibinfo{publisher}{Association for Computing Machinery},
  \bibinfo{address}{New York, NY, USA}, \bibinfo{pages}{159--166}.
\newblock
\urldef\tempurl%
\url{https://dl.acm.org/doi/10.1145/302979.303030}
\showURL{%
\tempurl}


\bibitem[Hui and Yu(2005)]%
        {hui2005extracting}
\bibfield{author}{\bibinfo{person}{Bowen Hui} {and} \bibinfo{person}{Eric Yu}.}
  \bibinfo{year}{2005}\natexlab{}.
\newblock \showarticletitle{Extracting conceptual relationships from
  specialized documents}.
\newblock \bibinfo{journal}{\emph{Data \& Knowledge Engineering}}
  \bibinfo{volume}{54}, \bibinfo{number}{1} (\bibinfo{year}{2005}),
  \bibinfo{pages}{29--55}.
\newblock


\bibitem[Jahanbakhsh et~al\mbox{.}(2022)]%
        {jahanbakhsh_understanding_2022}
\bibfield{author}{\bibinfo{person}{Farnaz Jahanbakhsh}, \bibinfo{person}{Elnaz
  Nouri}, \bibinfo{person}{Robert Sim}, \bibinfo{person}{Ryen~W. White}, {and}
  \bibinfo{person}{Adam Fourney}.} \bibinfo{year}{2022}\natexlab{}.
\newblock \showarticletitle{Understanding Questions that Arise When Working
  with Business Documents}.
\newblock \bibinfo{journal}{\emph{Proceedings of the ACM on Human-Computer
  Interaction}} \bibinfo{volume}{6}, \bibinfo{number}{CSCW2}
  (\bibinfo{date}{Nov.} \bibinfo{year}{2022}), \bibinfo{pages}{1--24}.
\newblock
\urldef\tempurl%
\url{https://dl.acm.org/doi/10.1145/3555761}
\showURL{%
\tempurl}


\bibitem[Ji et~al\mbox{.}(2023)]%
        {ji_survey_2023}
\bibfield{author}{\bibinfo{person}{Ziwei Ji}, \bibinfo{person}{Nayeon Lee},
  \bibinfo{person}{Rita Frieske}, \bibinfo{person}{Tiezheng Yu},
  \bibinfo{person}{Dan Su}, \bibinfo{person}{Yan Xu}, \bibinfo{person}{Etsuko
  Ishii}, \bibinfo{person}{Ye~Jin Bang}, \bibinfo{person}{Andrea Madotto},
  {and} \bibinfo{person}{Pascale Fung}.} \bibinfo{year}{2023}\natexlab{}.
\newblock \showarticletitle{Survey of Hallucination in Natural Language
  Generation}.
\newblock \bibinfo{journal}{\emph{ACM Comput. Surv.}} \bibinfo{volume}{55},
  \bibinfo{number}{12}, Article \bibinfo{articleno}{248} (\bibinfo{date}{March}
  \bibinfo{year}{2023}), \bibinfo{numpages}{38}~pages.
\newblock
\urldef\tempurl%
\url{https://doi.org/10.1145/3571730}
\showURL{%
\tempurl}


\bibitem[Jiang et~al\mbox{.}(2023)]%
        {jiang_graphologue_2023}
\bibfield{author}{\bibinfo{person}{Peiling Jiang}, \bibinfo{person}{Jude
  Rayan}, \bibinfo{person}{Steven~P. Dow}, {and} \bibinfo{person}{Haijun Xia}.}
  \bibinfo{year}{2023}\natexlab{}.
\newblock \showarticletitle{Graphologue: Exploring Large Language Model
  Responses with Interactive Diagrams}. In
  \bibinfo{booktitle}{\emph{Proceedings of the 36th Annual ACM Symposium on
  User Interface Software and Technology}} (San Francisco, CA, USA)
  \emph{(\bibinfo{series}{UIST '23})}. \bibinfo{publisher}{Association for
  Computing Machinery}, \bibinfo{address}{New York, NY, USA}, Article
  \bibinfo{articleno}{3}, \bibinfo{numpages}{20}~pages.
\newblock
\urldef\tempurl%
\url{https://doi.org/10.1145/3586183.3606737}
\showURL{%
\tempurl}


\bibitem[Jupyter(2023)]%
        {jupyter}
\bibfield{author}{\bibinfo{person}{Jupyter}.} \bibinfo{year}{2023}\natexlab{}.
\newblock \bibinfo{title}{Project Jupyter | Home}.
\newblock
\newblock
\urldef\tempurl%
\url{https://jupyter.org/}
\showURL{%
Retrieved September 1, 2023 from \tempurl}


\bibitem[Kang et~al\mbox{.}(2022)]%
        {kang_threddy_2022}
\bibfield{author}{\bibinfo{person}{Hyeonsu Kang}, \bibinfo{person}{Joseph~Chee
  Chang}, \bibinfo{person}{Yongsung Kim}, {and} \bibinfo{person}{Aniket
  Kittur}.} \bibinfo{year}{2022}\natexlab{}.
\newblock \showarticletitle{Threddy: An Interactive System for Personalized
  Thread-based Exploration and Organization of Scientific Literature}. In
  \bibinfo{booktitle}{\emph{Proceedings of the 35th Annual ACM Symposium on
  User Interface Software and Technology}} \emph{(\bibinfo{series}{UIST '22})}.
  \bibinfo{publisher}{Association for Computing Machinery},
  \bibinfo{address}{New York, NY, USA}, \bibinfo{pages}{1--15}.
\newblock
\urldef\tempurl%
\url{https://dl.acm.org/doi/10.1145/3526113.3545660}
\showURL{%
\tempurl}


\bibitem[Kang et~al\mbox{.}(2023)]%
        {kang_synergi_2023}
\bibfield{author}{\bibinfo{person}{Hyeonsu~B Kang}, \bibinfo{person}{Tongshuang
  Wu}, \bibinfo{person}{Joseph~Chee Chang}, {and} \bibinfo{person}{Aniket
  Kittur}.} \bibinfo{year}{2023}\natexlab{}.
\newblock \showarticletitle{Synergi: A Mixed-Initiative System for Scholarly
  Synthesis and Sensemaking}. In \bibinfo{booktitle}{\emph{Proceedings of the
  36th Annual ACM Symposium on User Interface Software and Technology}} (San
  Francisco, CA, USA) \emph{(\bibinfo{series}{UIST '23})}.
  \bibinfo{publisher}{Association for Computing Machinery},
  \bibinfo{address}{New York, NY, USA}, Article \bibinfo{articleno}{43},
  \bibinfo{numpages}{19}~pages.
\newblock
\urldef\tempurl%
\url{https://doi.org/10.1145/3586183.3606759}
\showURL{%
\tempurl}


\bibitem[Khot et~al\mbox{.}(2023)]%
        {khot2023decomposed}
\bibfield{author}{\bibinfo{person}{Tushar Khot}, \bibinfo{person}{Harsh
  Trivedi}, \bibinfo{person}{Matthew Finlayson}, \bibinfo{person}{Yao Fu},
  \bibinfo{person}{Kyle Richardson}, \bibinfo{person}{Peter Clark}, {and}
  \bibinfo{person}{Ashish Sabharwal}.} \bibinfo{year}{2023}\natexlab{}.
\newblock \bibinfo{title}{Decomposed Prompting: A Modular Approach for Solving
  Complex Tasks}.
\newblock
\newblock
\showeprint[arxiv]{2210.02406}~[cs.CL]


\bibitem[Kittur et~al\mbox{.}(2013)]%
        {kittur_costs_2013}
\bibfield{author}{\bibinfo{person}{Aniket Kittur}, \bibinfo{person}{Andrew~M.
  Peters}, \bibinfo{person}{Abdigani Diriye}, \bibinfo{person}{Trupti Telang},
  {and} \bibinfo{person}{Michael~R. Bove}.} \bibinfo{year}{2013}\natexlab{}.
\newblock \showarticletitle{Costs and benefits of structured information
  foraging}. In \bibinfo{booktitle}{\emph{Proceedings of the SIGCHI Conference
  on Human Factors in Computing Systems}}. \bibinfo{publisher}{ACM},
  \bibinfo{address}{Paris France}, \bibinfo{pages}{2989--2998}.
\newblock
\urldef\tempurl%
\url{https://dl.acm.org/doi/10.1145/2470654.2481415}
\showURL{%
\tempurl}


\bibitem[Kocielnik et~al\mbox{.}(2019)]%
        {kocielnik2019will}
\bibfield{author}{\bibinfo{person}{Rafal Kocielnik}, \bibinfo{person}{Saleema
  Amershi}, {and} \bibinfo{person}{Paul~N. Bennett}.}
  \bibinfo{year}{2019}\natexlab{}.
\newblock \showarticletitle{Will You Accept an Imperfect AI? Exploring Designs
  for Adjusting End-user Expectations of AI Systems}. In
  \bibinfo{booktitle}{\emph{Proceedings of the 2019 CHI Conference on Human
  Factors in Computing Systems}} \emph{(\bibinfo{series}{CHI '19})}.
  \bibinfo{publisher}{Association for Computing Machinery},
  \bibinfo{address}{New York, NY, USA}, \bibinfo{pages}{1--14}.
\newblock
\urldef\tempurl%
\url{https://dl.acm.org/doi/10.1145/3290605.3300641}
\showURL{%
\tempurl}


\bibitem[Kuznetsov et~al\mbox{.}(2022)]%
        {kuznetsov_fuse_2022}
\bibfield{author}{\bibinfo{person}{Andrew Kuznetsov},
  \bibinfo{person}{Joseph~Chee Chang}, \bibinfo{person}{Nathan Hahn},
  \bibinfo{person}{Napol Rachatasumrit}, \bibinfo{person}{Bradley Breneisen},
  \bibinfo{person}{Julina Coupland}, {and} \bibinfo{person}{Aniket Kittur}.}
  \bibinfo{year}{2022}\natexlab{}.
\newblock \showarticletitle{Fuse: In-Situ Sensemaking Support in the Browser}.
  In \bibinfo{booktitle}{\emph{Proceedings of the 35th Annual ACM Symposium on
  User Interface Software and Technology}} \emph{(\bibinfo{series}{UIST '22})}.
  \bibinfo{publisher}{Association for Computing Machinery},
  \bibinfo{address}{New York, NY, USA}, \bibinfo{pages}{1--15}.
\newblock
\urldef\tempurl%
\url{https://dl.acm.org/doi/10.1145/3526113.3545693}
\showURL{%
\tempurl}


\bibitem[Liu et~al\mbox{.}(2019)]%
        {liu_unakite_2019}
\bibfield{author}{\bibinfo{person}{Michael~Xieyang Liu}, \bibinfo{person}{Jane
  Hsieh}, \bibinfo{person}{Nathan Hahn}, \bibinfo{person}{Angelina Zhou},
  \bibinfo{person}{Emily Deng}, \bibinfo{person}{Shaun Burley},
  \bibinfo{person}{Cynthia Taylor}, \bibinfo{person}{Aniket Kittur}, {and}
  \bibinfo{person}{Brad~A. Myers}.} \bibinfo{year}{2019}\natexlab{}.
\newblock \showarticletitle{Unakite: Scaffolding Developers' Decision-Making
  Using the Web}. In \bibinfo{booktitle}{\emph{Proceedings of the 32nd Annual
  ACM Symposium on User Interface Software and Technology}}.
  \bibinfo{publisher}{ACM}, \bibinfo{address}{New Orleans LA USA},
  \bibinfo{pages}{67--80}.
\newblock
\urldef\tempurl%
\url{https://dl.acm.org/doi/10.1145/3332165.3347908}
\showURL{%
\tempurl}


\bibitem[Liu et~al\mbox{.}(2021)]%
        {liu_reuse_2021}
\bibfield{author}{\bibinfo{person}{Michael~Xieyang Liu},
  \bibinfo{person}{Aniket Kittur}, {and} \bibinfo{person}{Brad~A. Myers}.}
  \bibinfo{year}{2021}\natexlab{}.
\newblock \showarticletitle{To Reuse or Not To Reuse?: A Framework and System
  for Evaluating Summarized Knowledge}.
\newblock \bibinfo{journal}{\emph{Proceedings of the ACM on Human-Computer
  Interaction}} \bibinfo{volume}{5}, \bibinfo{number}{CSCW1}
  (\bibinfo{date}{April} \bibinfo{year}{2021}), \bibinfo{pages}{1--35}.
\newblock
\urldef\tempurl%
\url{https://dl.acm.org/doi/10.1145/3449240}
\showURL{%
\tempurl}


\bibitem[Liu et~al\mbox{.}(2022a)]%
        {liu_crystalline_2022}
\bibfield{author}{\bibinfo{person}{Michael~Xieyang Liu},
  \bibinfo{person}{Aniket Kittur}, {and} \bibinfo{person}{Brad~A. Myers}.}
  \bibinfo{year}{2022}\natexlab{a}.
\newblock \showarticletitle{Crystalline: Lowering the Cost for Developers to
  Collect and Organize Information for Decision Making}. In
  \bibinfo{booktitle}{\emph{Proceedings of the 2022 CHI Conference on Human
  Factors in Computing Systems}} \emph{(\bibinfo{series}{CHI '22})}.
  \bibinfo{publisher}{Association for Computing Machinery},
  \bibinfo{address}{New York, NY, USA}, \bibinfo{pages}{1--16}.
\newblock
\urldef\tempurl%
\url{https://dl.acm.org/doi/10.1145/3491102.3501968}
\showURL{%
\tempurl}


\bibitem[Liu et~al\mbox{.}(2022b)]%
        {liu_wigglite_2022}
\bibfield{author}{\bibinfo{person}{Michael~Xieyang Liu},
  \bibinfo{person}{Andrew Kuznetsov}, \bibinfo{person}{Yongsung Kim},
  \bibinfo{person}{Joseph~Chee Chang}, \bibinfo{person}{Aniket Kittur}, {and}
  \bibinfo{person}{Brad~A. Myers}.} \bibinfo{year}{2022}\natexlab{b}.
\newblock \showarticletitle{Wigglite: Low-Cost Information Collection and
  Triage}. In \bibinfo{booktitle}{\emph{Proceedings of the 35th Annual ACM
  Symposium on User Interface Software and Technology}} (Bend, OR, USA)
  \emph{(\bibinfo{series}{UIST '22})}. \bibinfo{publisher}{Association for
  Computing Machinery}, \bibinfo{address}{New York, NY, USA}, Article
  \bibinfo{articleno}{32}, \bibinfo{numpages}{16}~pages.
\newblock
\urldef\tempurl%
\url{https://doi.org/10.1145/3526113.3545661}
\showURL{%
\tempurl}


\bibitem[Lo et~al\mbox{.}(2023)]%
        {lo2023semantic}
\bibfield{author}{\bibinfo{person}{Kyle Lo}, \bibinfo{person}{Joseph~Chee
  Chang}, \bibinfo{person}{Andrew Head}, \bibinfo{person}{Jonathan Bragg},
  \bibinfo{person}{Amy~X. Zhang}, \bibinfo{person}{Cassidy Trier},
  \bibinfo{person}{Chloe Anastasiades}, \bibinfo{person}{Tal August},
  \bibinfo{person}{Russell Authur}, \bibinfo{person}{Danielle Bragg},
  \bibinfo{person}{Erin Bransom}, \bibinfo{person}{Isabel Cachola},
  \bibinfo{person}{Stefan Candra}, \bibinfo{person}{Yoganand Chandrasekhar},
  \bibinfo{person}{Yen-Sung Chen}, \bibinfo{person}{Evie Yu-Yen Cheng},
  \bibinfo{person}{Yvonne Chou}, \bibinfo{person}{Doug Downey},
  \bibinfo{person}{Rob Evans}, \bibinfo{person}{Raymond Fok},
  \bibinfo{person}{Fangzhou Hu}, \bibinfo{person}{Regan Huff},
  \bibinfo{person}{Dongyeop Kang}, \bibinfo{person}{Tae~Soo Kim},
  \bibinfo{person}{Rodney Kinney}, \bibinfo{person}{Aniket Kittur},
  \bibinfo{person}{Hyeonsu Kang}, \bibinfo{person}{Egor Klevak},
  \bibinfo{person}{Bailey Kuehl}, \bibinfo{person}{Michael Langan},
  \bibinfo{person}{Matt Latzke}, \bibinfo{person}{Jaron Lochner},
  \bibinfo{person}{Kelsey MacMillan}, \bibinfo{person}{Eric Marsh},
  \bibinfo{person}{Tyler Murray}, \bibinfo{person}{Aakanksha Naik},
  \bibinfo{person}{Ngoc-Uyen Nguyen}, \bibinfo{person}{Srishti Palani},
  \bibinfo{person}{Soya Park}, \bibinfo{person}{Caroline Paulic},
  \bibinfo{person}{Napol Rachatasumrit}, \bibinfo{person}{Smita Rao},
  \bibinfo{person}{Paul Sayre}, \bibinfo{person}{Zejiang Shen},
  \bibinfo{person}{Pao Siangliulue}, \bibinfo{person}{Luca Soldaini},
  \bibinfo{person}{Huy Tran}, \bibinfo{person}{Madeleine van Zuylen},
  \bibinfo{person}{Lucy~Lu Wang}, \bibinfo{person}{Christopher Wilhelm},
  \bibinfo{person}{Caroline Wu}, \bibinfo{person}{Jiangjiang Yang},
  \bibinfo{person}{Angele Zamarron}, \bibinfo{person}{Marti~A. Hearst}, {and}
  \bibinfo{person}{Daniel~S. Weld}.} \bibinfo{year}{2023}\natexlab{}.
\newblock \bibinfo{title}{The Semantic Reader Project: Augmenting Scholarly
  Documents through AI-Powered Interactive Reading Interfaces}.
\newblock
\newblock
\showeprint[arxiv]{2303.14334}~[cs.HC]


\bibitem[Matejka et~al\mbox{.}(2021)]%
        {matejka2021paperForager}
\bibfield{author}{\bibinfo{person}{Justin Matejka}, \bibinfo{person}{Tovi
  Grossman}, {and} \bibinfo{person}{George Fitzmaurice}.}
  \bibinfo{year}{2021}\natexlab{}.
\newblock \showarticletitle{Paper Forager: Supporting the Rapid Exploration of
  Research Document Collections}. In \bibinfo{booktitle}{\emph{Proceedings of
  Graphics Interface 2021}} \emph{(\bibinfo{series}{GI 2021})}.
  \bibinfo{publisher}{Canadian Information Processing Society},
  \bibinfo{address}{Virtual Event}, \bibinfo{pages}{237 -- 245}.
\newblock


\bibitem[Moretti et~al\mbox{.}(2016)]%
        {moretti2016alcide}
\bibfield{author}{\bibinfo{person}{Giovanni Moretti}, \bibinfo{person}{Rachele
  Sprugnoli}, \bibinfo{person}{Stefano Menini}, {and} \bibinfo{person}{Sara
  Tonelli}.} \bibinfo{year}{2016}\natexlab{}.
\newblock \showarticletitle{ALCIDE: Extracting and visualising content from
  large document collections to support humanities studies}.
\newblock \bibinfo{journal}{\emph{Knowledge-Based Systems}}
  \bibinfo{volume}{111} (\bibinfo{year}{2016}), \bibinfo{pages}{100--112}.
\newblock


\bibitem[Mündler et~al\mbox{.}(2023)]%
        {mundler_self-contradictory_2023}
\bibfield{author}{\bibinfo{person}{Niels Mündler}, \bibinfo{person}{Jingxuan
  He}, \bibinfo{person}{Slobodan Jenko}, {and} \bibinfo{person}{Martin
  Vechev}.} \bibinfo{year}{2023}\natexlab{}.
\newblock \bibinfo{title}{Self-contradictory Hallucinations of Large Language
  Models: Evaluation, Detection and Mitigation}.
\newblock
\newblock
\urldef\tempurl%
\url{http://arxiv.org/abs/2305.15852}
\showURL{%
\tempurl}


\bibitem[Nguyen et~al\mbox{.}(2016)]%
        {nguyen_sensemap_2016}
\bibfield{author}{\bibinfo{person}{Phong~H. Nguyen}, \bibinfo{person}{Kai Xu},
  \bibinfo{person}{Andy Bardill}, \bibinfo{person}{Betul Salman},
  \bibinfo{person}{Kate Herd}, {and} \bibinfo{person}{B.L.~William Wong}.}
  \bibinfo{year}{2016}\natexlab{}.
\newblock \showarticletitle{SenseMap: Supporting browser-based online
  sensemaking through analytic provenance}. In \bibinfo{booktitle}{\emph{2016
  IEEE Conference on Visual Analytics Science and Technology (VAST)}}.
  \bibinfo{publisher}{IEEE}, \bibinfo{address}{Baltimore, MD, USA},
  \bibinfo{pages}{91--100}.
\newblock


\bibitem[OpenAI(2023)]%
        {chatgpt}
\bibfield{author}{\bibinfo{person}{OpenAI}.} \bibinfo{year}{2023}\natexlab{}.
\newblock \bibinfo{title}{ChatGPT}.
\newblock
\newblock
\urldef\tempurl%
\url{https://chat.openai.com/}
\showURL{%
Retrieved December 4, 2023 from \tempurl}


\bibitem[Palani et~al\mbox{.}(2023)]%
        {palani_relatedly_2023}
\bibfield{author}{\bibinfo{person}{Srishti Palani}, \bibinfo{person}{Aakanksha
  Naik}, \bibinfo{person}{Doug Downey}, \bibinfo{person}{Amy~X. Zhang},
  \bibinfo{person}{Jonathan Bragg}, {and} \bibinfo{person}{Joseph~Chee Chang}.}
  \bibinfo{year}{2023}\natexlab{}.
\newblock \showarticletitle{Relatedly: Scaffolding Literature Reviews with
  Existing Related Work Sections}. In \bibinfo{booktitle}{\emph{Proceedings of
  the 2023 CHI Conference on Human Factors in Computing Systems}}.
  \bibinfo{publisher}{ACM}, \bibinfo{address}{Hamburg Germany},
  \bibinfo{pages}{1--20}.
\newblock
\urldef\tempurl%
\url{https://dl.acm.org/doi/10.1145/3544548.3580841}
\showURL{%
\tempurl}


\bibitem[Pirolli and Card(1995)]%
        {pirolli_information_1995}
\bibfield{author}{\bibinfo{person}{Peter Pirolli} {and} \bibinfo{person}{Stuart
  Card}.} \bibinfo{year}{1995}\natexlab{}.
\newblock \showarticletitle{Information foraging in information access
  environments}. In \bibinfo{booktitle}{\emph{Proceedings of the SIGCHI
  Conference on Human Factors in Computing Systems}}
  \emph{(\bibinfo{series}{CHI '95})}. \bibinfo{publisher}{ACM
  Press/Addison-Wesley Publishing Co.}, \bibinfo{address}{USA},
  \bibinfo{pages}{51--58}.
\newblock
\urldef\tempurl%
\url{https://dl.acm.org/doi/10.1145/223904.223911}
\showURL{%
\tempurl}


\bibitem[Pirolli and Card(2005)]%
        {pirolli_sensemaking_2005}
\bibfield{author}{\bibinfo{person}{Peter Pirolli} {and} \bibinfo{person}{Stuart
  Card}.} \bibinfo{year}{2005}\natexlab{}.
\newblock \showarticletitle{The sensemaking process and leverage points for
  analyst technology as identified through cognitive task analysis}. In
  \bibinfo{booktitle}{\emph{Proceedings of international conference on
  intelligence analysis}}, Vol.~\bibinfo{volume}{5}. \bibinfo{publisher}{IEEE},
  \bibinfo{address}{McLean, VA, USA}, \bibinfo{pages}{2--4}.
\newblock


\bibitem[Radhakrishnan et~al\mbox{.}(2023)]%
        {radhakrishnan2023question}
\bibfield{author}{\bibinfo{person}{Ansh Radhakrishnan}, \bibinfo{person}{Karina
  Nguyen}, \bibinfo{person}{Anna Chen}, \bibinfo{person}{Carol Chen},
  \bibinfo{person}{Carson Denison}, \bibinfo{person}{Danny Hernandez},
  \bibinfo{person}{Esin Durmus}, \bibinfo{person}{Evan Hubinger},
  \bibinfo{person}{Jackson Kernion}, \bibinfo{person}{Kamilė Lukošiūtė},
  \bibinfo{person}{Newton Cheng}, \bibinfo{person}{Nicholas Joseph},
  \bibinfo{person}{Nicholas Schiefer}, \bibinfo{person}{Oliver Rausch},
  \bibinfo{person}{Sam McCandlish}, \bibinfo{person}{Sheer~El Showk},
  \bibinfo{person}{Tamera Lanham}, \bibinfo{person}{Tim Maxwell},
  \bibinfo{person}{Venkatesa Chandrasekaran}, \bibinfo{person}{Zac
  Hatfield-Dodds}, \bibinfo{person}{Jared Kaplan}, \bibinfo{person}{Jan
  Brauner}, \bibinfo{person}{Samuel~R. Bowman}, {and} \bibinfo{person}{Ethan
  Perez}.} \bibinfo{year}{2023}\natexlab{}.
\newblock \bibinfo{title}{Question Decomposition Improves the Faithfulness of
  Model-Generated Reasoning}.
\newblock
\newblock
\showeprint[arxiv]{2307.11768}~[cs.CL]


\bibitem[Ramos et~al\mbox{.}(2022)]%
        {ramos_forsense_2022}
\bibfield{author}{\bibinfo{person}{Gonzalo Ramos}, \bibinfo{person}{Napol
  Rachatasumrit}, \bibinfo{person}{Jina Suh}, \bibinfo{person}{Rachel Ng},
  {and} \bibinfo{person}{Christopher Meek}.} \bibinfo{year}{2022}\natexlab{}.
\newblock \showarticletitle{ForSense: Accelerating Online Research Through
  Sensemaking Integration and Machine Research Support}.
\newblock \bibinfo{journal}{\emph{ACM Transactions on Interactive Intelligent
  Systems}} \bibinfo{volume}{12}, \bibinfo{number}{4} (\bibinfo{date}{Dec.}
  \bibinfo{year}{2022}), \bibinfo{pages}{1--23}.
\newblock
\urldef\tempurl%
\url{https://dl.acm.org/doi/10.1145/3532853}
\showURL{%
\tempurl}


\bibitem[Raudenbush and Bryk(2002)]%
        {raudenbush2002hierarchical}
\bibfield{author}{\bibinfo{person}{Stephen~W Raudenbush} {and}
  \bibinfo{person}{Anthony~S Bryk}.} \bibinfo{year}{2002}\natexlab{}.
\newblock \bibinfo{booktitle}{\emph{Hierarchical linear models: Applications
  and data analysis methods}}. Vol.~\bibinfo{volume}{1}.
\newblock \bibinfo{publisher}{SAGE}, \bibinfo{address}{Thousand Oaks, CA}.
\newblock


\bibitem[Reimers and Gurevych(2019)]%
        {reimers-2019-sentence-bert}
\bibfield{author}{\bibinfo{person}{Nils Reimers} {and} \bibinfo{person}{Iryna
  Gurevych}.} \bibinfo{year}{2019}\natexlab{}.
\newblock \showarticletitle{Sentence-BERT: Sentence Embeddings using Siamese
  BERT-Networks}. In \bibinfo{booktitle}{\emph{Proceedings of the 2019
  Conference on Empirical Methods in Natural Language Processing and the 9th
  International Joint Conference on Natural Language Processing
  (EMNLP-IJCNLP)}}. \bibinfo{publisher}{Association for Computational
  Linguistics}, \bibinfo{address}{Hong Kong, China},
  \bibinfo{pages}{3982--3992}.
\newblock
\urldef\tempurl%
\url{https://aclanthology.org/D19-1410}
\showURL{%
\tempurl}


\bibitem[Roegiest and Wei(2018)]%
        {roegiest2018redesigning}
\bibfield{author}{\bibinfo{person}{Adam Roegiest} {and} \bibinfo{person}{Winter
  Wei}.} \bibinfo{year}{2018}\natexlab{}.
\newblock \showarticletitle{Redesigning a Document Viewer for Legal Documents}.
  In \bibinfo{booktitle}{\emph{Proceedings of the 2018 Conference on Human
  Information Interaction \& Retrieval}} (New Brunswick, NJ, USA)
  \emph{(\bibinfo{series}{CHIIR '18})}. \bibinfo{publisher}{Association for
  Computing Machinery}, \bibinfo{address}{New York, NY, USA},
  \bibinfo{pages}{297–300}.
\newblock
\urldef\tempurl%
\url{https://doi.org/10.1145/3176349.3176873}
\showURL{%
\tempurl}


\bibitem[Russell et~al\mbox{.}(1993)]%
        {russell_cost_1993}
\bibfield{author}{\bibinfo{person}{Daniel~M. Russell}, \bibinfo{person}{Mark~J.
  Stefik}, \bibinfo{person}{Peter Pirolli}, {and} \bibinfo{person}{Stuart~K.
  Card}.} \bibinfo{year}{1993}\natexlab{}.
\newblock \showarticletitle{The cost structure of sensemaking}. In
  \bibinfo{booktitle}{\emph{Proceedings of the SIGCHI conference on Human
  factors in computing systems - CHI '93}}. \bibinfo{publisher}{ACM Press},
  \bibinfo{address}{Amsterdam, The Netherlands}, \bibinfo{pages}{269--276}.
\newblock
\urldef\tempurl%
\url{http://portal.acm.org/citation.cfm?doid=169059.169209}
\showURL{%
\tempurl}


\bibitem[schraefel et~al\mbox{.}(2006)]%
        {schraefel_mspace_2006}
\bibfield{author}{\bibinfo{person}{m.c. schraefel}, \bibinfo{person}{Max
  Wilson}, \bibinfo{person}{Alistair Russell}, {and} \bibinfo{person}{Daniel~A.
  Smith}.} \bibinfo{year}{2006}\natexlab{}.
\newblock \showarticletitle{mSpace: improving information access to multimedia
  domains with multimodal exploratory search}.
\newblock \bibinfo{journal}{\emph{Commun. ACM}} \bibinfo{volume}{49},
  \bibinfo{number}{4} (\bibinfo{date}{April} \bibinfo{year}{2006}),
  \bibinfo{pages}{47--49}.
\newblock
\urldef\tempurl%
\url{https://dl.acm.org/doi/10.1145/1121949.1121980}
\showURL{%
\tempurl}


\bibitem[schraefel et~al\mbox{.}(2002)]%
        {schraefel_hunter_2002}
\bibfield{author}{\bibinfo{person}{m.~c. schraefel}, \bibinfo{person}{Yuxiang
  Zhu}, \bibinfo{person}{David Modjeska}, \bibinfo{person}{Daniel Wigdor},
  {and} \bibinfo{person}{Shengdong Zhao}.} \bibinfo{year}{2002}\natexlab{}.
\newblock \showarticletitle{Hunter gatherer: interaction support for the
  creation and management of within-web-page collections}. In
  \bibinfo{booktitle}{\emph{Proceedings of the 11th international conference on
  World Wide Web}} \emph{(\bibinfo{series}{WWW '02})}.
  \bibinfo{publisher}{Association for Computing Machinery},
  \bibinfo{address}{New York, NY, USA}, \bibinfo{pages}{172--181}.
\newblock
\urldef\tempurl%
\url{https://dl.acm.org/doi/10.1145/511446.511469}
\showURL{%
\tempurl}


\bibitem[Services(2023)]%
        {opensearch}
\bibfield{author}{\bibinfo{person}{Amazon~Web Services}.}
  \bibinfo{year}{2023}\natexlab{}.
\newblock \bibinfo{title}{Using OpenSearch as a Vector Database}.
\newblock
\newblock
\urldef\tempurl%
\url{https://opensearch.org/platform/search/vector-database.html}
\showURL{%
Retrieved August 27, 2023 from \tempurl}


\bibitem[Shen et~al\mbox{.}(2023)]%
        {shen_chatgpt_2023}
\bibfield{author}{\bibinfo{person}{Yiqiu Shen}, \bibinfo{person}{Laura
  Heacock}, \bibinfo{person}{Jonathan Elias}, \bibinfo{person}{Keith~D.
  Hentel}, \bibinfo{person}{Beatriu Reig}, \bibinfo{person}{George Shih}, {and}
  \bibinfo{person}{Linda Moy}.} \bibinfo{year}{2023}\natexlab{}.
\newblock \showarticletitle{ChatGPT and Other Large Language Models Are
  Double-edged Swords}.
\newblock \bibinfo{journal}{\emph{Radiology}} \bibinfo{volume}{307},
  \bibinfo{number}{2} (\bibinfo{date}{April} \bibinfo{year}{2023}),
  \bibinfo{pages}{e230163}.
\newblock
\urldef\tempurl%
\url{https://pubs.rsna.org/doi/abs/10.1148/radiol.230163}
\showURL{%
\tempurl}


\bibitem[Shneiderman(1996)]%
        {eyes_schneiderman_1996}
\bibfield{author}{\bibinfo{person}{B. Shneiderman}.}
  \bibinfo{year}{1996}\natexlab{}.
\newblock \showarticletitle{The Eyes Have It: A Task by Data Type Taxonomy for
  Information Visualizations}. In \bibinfo{booktitle}{\emph{Visual Languages,
  IEEE Symposium on}}. \bibinfo{publisher}{IEEE Computer Society},
  \bibinfo{address}{Los Alamitos, CA, USA}, \bibinfo{pages}{336}.
\newblock
\showISSN{1049-2615}
\urldef\tempurl%
\url{https://doi.org/10.1109/VL.1996.545307}
\showDOI{\tempurl}


\bibitem[Siu et~al\mbox{.}(2022)]%
        {siu_supporting_2022}
\bibfield{author}{\bibinfo{person}{Alexa Siu}, \bibinfo{person}{Gene S-H~Kim},
  \bibinfo{person}{Sile O'Modhrain}, {and} \bibinfo{person}{Sean Follmer}.}
  \bibinfo{year}{2022}\natexlab{}.
\newblock \showarticletitle{Supporting Accessible Data Visualization Through
  Audio Data Narratives}. In \bibinfo{booktitle}{\emph{Proceedings of the 2022
  CHI Conference on Human Factors in Computing Systems}}
  \emph{(\bibinfo{series}{CHI '22})}. \bibinfo{publisher}{Association for
  Computing Machinery}, \bibinfo{address}{New York, NY, USA},
  \bibinfo{pages}{1--19}.
\newblock
\urldef\tempurl%
\url{https://doi.org/10.1145/3491102.3517678}
\showURL{%
\tempurl}


\bibitem[Source(2023)]%
        {react}
\bibfield{author}{\bibinfo{person}{Meta~Open Source}.}
  \bibinfo{year}{2023}\natexlab{}.
\newblock \bibinfo{title}{React}.
\newblock
\newblock
\urldef\tempurl%
\url{https://react.dev/}
\showURL{%
Retrieved August 27, 2023 from \tempurl}


\bibitem[Suh et~al\mbox{.}(2023)]%
        {suh_sensecape_2023}
\bibfield{author}{\bibinfo{person}{Sangho Suh}, \bibinfo{person}{Bryan Min},
  \bibinfo{person}{Srishti Palani}, {and} \bibinfo{person}{Haijun Xia}.}
  \bibinfo{year}{2023}\natexlab{}.
\newblock \showarticletitle{Sensecape: Enabling Multilevel Exploration and
  Sensemaking with Large Language Models}. In
  \bibinfo{booktitle}{\emph{Proceedings of the 36th Annual ACM Symposium on
  User Interface Software and Technology}} (San Francisco, CA, USA)
  \emph{(\bibinfo{series}{UIST '23})}. \bibinfo{publisher}{Association for
  Computing Machinery}, \bibinfo{address}{New York, NY, USA}, Article
  \bibinfo{articleno}{1}, \bibinfo{numpages}{18}~pages.
\newblock
\urldef\tempurl%
\url{https://doi.org/10.1145/3586183.3606756}
\showURL{%
\tempurl}


\bibitem[Swearngin et~al\mbox{.}(2021)]%
        {swearngin_scraps_2021}
\bibfield{author}{\bibinfo{person}{Amanda Swearngin}, \bibinfo{person}{Shamsi
  Iqbal}, \bibinfo{person}{Victor Poznanski}, \bibinfo{person}{Mark
  Encarnación}, \bibinfo{person}{Paul~N. Bennett}, {and}
  \bibinfo{person}{Jaime Teevan}.} \bibinfo{year}{2021}\natexlab{}.
\newblock \showarticletitle{Scraps: Enabling Mobile Capture, Contextualization,
  and Use of Document Resources}. In \bibinfo{booktitle}{\emph{Proceedings of
  the 2021 CHI Conference on Human Factors in Computing Systems}}.
  \bibinfo{publisher}{ACM}, \bibinfo{address}{Yokohama Japan},
  \bibinfo{pages}{1--14}.
\newblock
\urldef\tempurl%
\url{https://dl.acm.org/doi/10.1145/3411764.3445185}
\showURL{%
\tempurl}


\bibitem[ter Hoeve et~al\mbox{.}(2020)]%
        {ter_hoeve_conversations_2020}
\bibfield{author}{\bibinfo{person}{Maartje ter Hoeve}, \bibinfo{person}{Robert
  Sim}, \bibinfo{person}{Elnaz Nouri}, \bibinfo{person}{Adam Fourney},
  \bibinfo{person}{Maarten de Rijke}, {and} \bibinfo{person}{Ryen~W. White}.}
  \bibinfo{year}{2020}\natexlab{}.
\newblock \showarticletitle{Conversations with Documents: An Exploration of
  Document-Centered Assistance}. In \bibinfo{booktitle}{\emph{Proceedings of
  the 2020 Conference on Human Information Interaction and Retrieval}}
  (Vancouver BC, Canada) \emph{(\bibinfo{series}{CHIIR '20})}.
  \bibinfo{publisher}{Association for Computing Machinery},
  \bibinfo{address}{New York, NY, USA}, \bibinfo{pages}{43–52}.
\newblock
\urldef\tempurl%
\url{https://doi.org/10.1145/3343413.3377971}
\showURL{%
\tempurl}


\bibitem[Vuong(1989)]%
        {vuong1989likelihood}
\bibfield{author}{\bibinfo{person}{Quang~H. Vuong}.}
  \bibinfo{year}{1989}\natexlab{}.
\newblock \showarticletitle{Likelihood Ratio Tests for Model Selection and
  Non-Nested Hypotheses}.
\newblock \bibinfo{journal}{\emph{Econometrica}} \bibinfo{volume}{57},
  \bibinfo{number}{2} (\bibinfo{year}{1989}), \bibinfo{pages}{307--333}.
\newblock
\urldef\tempurl%
\url{http://www.jstor.org/stable/1912557}
\showURL{%
\tempurl}


\bibitem[Wu et~al\mbox{.}(2023)]%
        {wu2023multimodal}
\bibfield{author}{\bibinfo{person}{Jiayang Wu}, \bibinfo{person}{Wensheng Gan},
  \bibinfo{person}{Zefeng Chen}, \bibinfo{person}{Shicheng Wan}, {and}
  \bibinfo{person}{Philip~S. Yu}.} \bibinfo{year}{2023}\natexlab{}.
\newblock \bibinfo{title}{Multimodal Large Language Models: A Survey}.
\newblock
\newblock
\showeprint[arxiv]{2311.13165}~[cs.AI]


\bibitem[Zapier(2021)]%
        {zapier_office}
\bibfield{author}{\bibinfo{person}{Zapier}.} \bibinfo{year}{2021}\natexlab{}.
\newblock \bibinfo{title}{Zapier report: {How} office workers spend their
  time}.
\newblock
\newblock
\urldef\tempurl%
\url{https://zapier.com/blog/report-how-office-workers-spend-time/}
\showURL{%
\tempurl}


\bibitem[Zhang et~al\mbox{.}(2008)]%
        {zhang_citesense_2008}
\bibfield{author}{\bibinfo{person}{Xiaolong Zhang}, \bibinfo{person}{Yan Qu},
  \bibinfo{person}{C.~Lee Giles}, {and} \bibinfo{person}{Piyou Song}.}
  \bibinfo{year}{2008}\natexlab{}.
\newblock \showarticletitle{CiteSense: supporting sensemaking of research
  literature}. In \bibinfo{booktitle}{\emph{Proceedings of the SIGCHI
  Conference on Human Factors in Computing Systems}}
  \emph{(\bibinfo{series}{CHI '08})}. \bibinfo{publisher}{Association for
  Computing Machinery}, \bibinfo{address}{New York, NY, USA},
  \bibinfo{pages}{677--680}.
\newblock
\urldef\tempurl%
\url{https://dl.acm.org/doi/10.1145/1357054.1357161}
\showURL{%
\tempurl}


\bibitem[Zhang et~al\mbox{.}(2023)]%
        {zhang_visar_2023}
\bibfield{author}{\bibinfo{person}{Zheng Zhang}, \bibinfo{person}{Jie Gao},
  \bibinfo{person}{Ranjodh~Singh Dhaliwal}, {and} \bibinfo{person}{Toby Jia-Jun
  Li}.} \bibinfo{year}{2023}\natexlab{}.
\newblock \showarticletitle{VISAR: A Human-AI Argumentative Writing Assistant
  with Visual Programming and Rapid Draft Prototyping}. In
  \bibinfo{booktitle}{\emph{Proceedings of the 36th Annual ACM Symposium on
  User Interface Software and Technology}} (San Francisco, CA, USA)
  \emph{(\bibinfo{series}{UIST '23})}. \bibinfo{publisher}{Association for
  Computing Machinery}, \bibinfo{address}{New York, NY, USA}, Article
  \bibinfo{articleno}{5}, \bibinfo{numpages}{30}~pages.
\newblock
\urldef\tempurl%
\url{https://doi.org/10.1145/3586183.3606800}
\showURL{%
\tempurl}


\bibitem[Zhao and Lee(2020)]%
        {zhao_talk_2020}
\bibfield{author}{\bibinfo{person}{Tiancheng Zhao} {and}
  \bibinfo{person}{Kyusong Lee}.} \bibinfo{year}{2020}\natexlab{}.
\newblock \showarticletitle{Talk to Papers: Bringing Neural Question Answering
  to Academic Search}. In \bibinfo{booktitle}{\emph{Proceedings of the 58th
  Annual Meeting of the Association for Computational Linguistics: System
  Demonstrations}}. \bibinfo{publisher}{Association for Computational
  Linguistics}, \bibinfo{address}{Online}, \bibinfo{pages}{30--36}.
\newblock
\urldef\tempurl%
\url{https://aclanthology.org/2020.acl-demos.5}
\showURL{%
\tempurl}


\end{thebibliography}
